\documentclass[floatfix, a4paper, amsfonts, amssymb, amsmath, reprint, showkeys, nofootinbib, twoside, superscriptaddress, aps, prl]{revtex4-2}
\usepackage{filecontents}	
\usepackage{graphicx}
\usepackage{physics}
\usepackage{mathtools}
\usepackage{colonequals}
\usepackage[caption=false]{subfig}
\usepackage{nicefrac}
\usepackage{color, colortbl}
\usepackage{multirow}
\usepackage[normalem]{ulem} 
\usepackage{float}
\usepackage{xr}

\newcommand{\ra}{\rightarrow}

\newcommand{\Vsup}{V_\mathrm{S}}
\newcommand{\Vmode}{V_\mathrm{M}}
\newcommand{\ie}{\textit{i.e.}}
\newcommand{\eg}{\textit{e.g.}}

\renewcommand{\dd}{\mathrm{d}}
\newcommand{\Vcav}{V_\mathrm{C}}
\newcommand{\Ecav}{E_\mathrm{C}}
\newcommand{\Esup}{E_\mathrm{S}}
\newcommand{\const}{\mathrm{const}}

\newcommand{\eq}[1]{(\ref{#1})}

\newcommand{\VV}{\widetilde V}
\newcommand{\EE}{\widetilde E}

\usepackage[T1]{fontenc}
\usepackage{notes2bib}
\bibnotesetup{
  note-name = ,
  use-sort-key = false
}
\begin{document}
\title{Scaling theory of wave confinement in classical and quantum periodic systems}
\author{Marek Kozo\v{n}}
\affiliation{Complex Photonic Systems (COPS), MESA+ Institute for Nanotechnology, University of Twente, P.O. Box 217, 7500 AE Enschede, The Netherlands}
\affiliation{Mathematics of Computational Science (MACS),  MESA+ Institute for Nanotechnology, University of Twente, P.O. Box 217, 7500 AE Enschede, The Netherlands}

\author{Ad Lagendijk}
\affiliation{Complex Photonic Systems (COPS), MESA+ Institute for Nanotechnology, University of Twente, P.O. Box 217, 7500 AE Enschede, The Netherlands}

\author{Matthias Schlottbom}
\affiliation{Mathematics of Computational Science (MACS),  MESA+ Institute for Nanotechnology, University of Twente, P.O. Box 217, 7500 AE Enschede, The Netherlands}

\author{Jaap J. W. van der Vegt}
\affiliation{Mathematics of Computational Science (MACS),  MESA+ Institute for Nanotechnology, University of Twente, P.O. Box 217, 7500 AE Enschede, The Netherlands}

\author{Willem L. Vos}
\affiliation{Complex Photonic Systems (COPS), MESA+ Institute for Nanotechnology, University of Twente, P.O. Box 217, 7500 AE Enschede, The Netherlands}

\begin{abstract}
Functional defects in periodic media confine waves - acoustic, electromagnetic, electronic, spin, \textit{etc}. - in various dimensions, depending on the structure of the defect. 
While defects are usually modelled by a superlattice with a typical band-structure representation of energy levels, determining the confinement associated with a given band is highly non-trivial and no analytical method is known to date. 
Therefore, we propose a rigorous method to classify the dimensionality of the confinement. 
Starting from the confinement energy and the mode volume, we use finite-size scaling to find that ratios of these quantities to certain powers yield the confinement dimensionality of each band. 
This classification has negligible additional computational costs compared to a band structure calculation and is valid for any type of wave in both quantum and classical regimes, and any dimension. 
In the quantum case, we illustrate our method on electronic confinement in 2D hexagonal BN with a nitrogen vacancy, which confirms the previous results. 
In the classical case, we study a three-dimensional photonic band gap cavity superlattice, where we identify novel acceptor-like behavior. 
\end{abstract}
\maketitle

Completely controlling the propagation of waves in periodic media is a key challenge that is essential for a large variety of applications ~\cite{Markos2008, Fink2000Rep.Prog.Phys., Liu2000Science, Maldovan2013Nature, Cummer2016Nat.Rev.Mater., Kruglyak2010J.Phys.D, Wagner2016Nat.Nanotechnol., Klyukin2018Phys.Rev.Lett., Callahan2013Opt.Express, Tandaechanurat2011Nat.Photonics, Aspelmeyer2014Rev.Mod.Phys., Koenderink2015Science, Li2018Opt.Express, Wang2020Nat.Photonics, Uppu2021Phys.Rev.Lett.}. 
An especially interesting type of propagation control is wave confinement achieved by introducing disorder and functional defects into an otherwise periodic medium. 
The interference of waves in such an altered structure may result in a strong concentration of the energy density inside a small sub-volume of the medium.
Wave confinement has been investigated for different types of waves and in various settings, for example, classical mechanics~\cite{Arceri2020Phys.Rev.Lett.}, photonics~\cite{Busch2007Phys.Rep., Woldering2014Phys.Rev.B, Hack2019Phys.Rev.B, Callahan2013Opt.Express}, solid state physics~\cite{Economou2010,Shao2008J.Phys.Chem.C, Pashartis2017Phys.Rev.Appl., Pashartis2017Phys.Rev.B, Zhang2011Phys.Rev.B}, or magnonics~\cite{Demokritov2017book, Tartakovskaya2016Phys.Rev.B}.
Applications of this phenomenon include sensors, enhanced spontaneous emission, and enhanced interactions between hybrid wave-types such as sound and light~\cite{Krioukov2002Opt.Lett.,Baba2008Nat.Photonics, Noda2000Nature,Gerard1998Phys.Rev.Lett.,Michler2003,Reithmaier2004Nature,Yoshie2004Nature,Peter2005Phys.Rev.Lett.,Russell2003Opt.Express}.

The analysis of spatial concentration of energy in physical systems is in photonics traditionally done via the mode volume~\cite{Woldering2014Phys.Rev.B, Painter1999Science, Sauvan2013Phys.Rev.Lett., Kristensen2015Phys.Rev.A, Muljarov2016Phys.Rev.B}  or equivalently, in condensed matter physics, via the participation ratio~\cite{El-Dardiry2012Phys.Rev.B, Arceri2020Phys.Rev.Lett., Pashartis2017Phys.Rev.B}. 
Bands with small mode volume or participation ratio are considered to be confined while, conversely, bands with large mode volume or participation ratio are taken to be extended~\cite{Vahala2003Nature, Arceri2020Phys.Rev.Lett., Woldering2014Phys.Rev.B, Hack2019Phys.Rev.B, Pashartis2017Phys.Rev.Appl., Pashartis2017Phys.Rev.B, Zhang2011Phys.Rev.B}.
However, the notion of what specifically is 'large' and 'small' is in each case determined rather subjectively and there are no rigorous boundaries imposed by nature.

An alternative method to analyze the wave confinement, called multifractality analysis~\cite{Aoki1983J.Phys.C:SolidStatePhys., Janssen1998Phys.Rep., Faez2009Phys.Rev.Lett.}, is based on the participation ratio scaling in the limit of infinitely large supercells. Unfortunately, this approach requires impractically large supercells to yield reliable results, further compounded by its inability to deal with band folding (see Supplemental material).

In this Letter, we present a rigorous method to determine confinement of waves in periodic structures with defects, based on finite-size scaling~\cite{Abrahams1979Phys.Rev.Lett., Sheng2006book, Slevin2001Phys.Rev.Lett.}.
Rather than extending the supercells towards infinity, our method determines confinement by looking at supercells of finite size.
As a consequence, our approach requires only minimal computational overhead and its results are directly relevant for experimentally interesting finitely-large systems.
This technique can be viewed as a more accessible and practical extension of the multifractality concept, also suitable for automated classification.

Superimposing a defect lattice on an unperturbed crystal lattice gives rise to a defect superlattice~\cite{Pashartis2017Phys.Rev.Appl., Pashartis2017Phys.Rev.B,Kruglyak2010J.Phys.D,Bragg1934Proc.R.Soc.Lond.A,Bethe1935Proc.R.Soc.Lond.A, Grahn1995book}.
A unit cell of such a superlattice, containing $N^D$ unit cells, where $D$ is the system dimension, will be referred to as a supercell of linear size $N$.
For simplicity, we keep $N$ uniform for all directions.

\begin{figure}
{\includegraphics[width=\columnwidth]{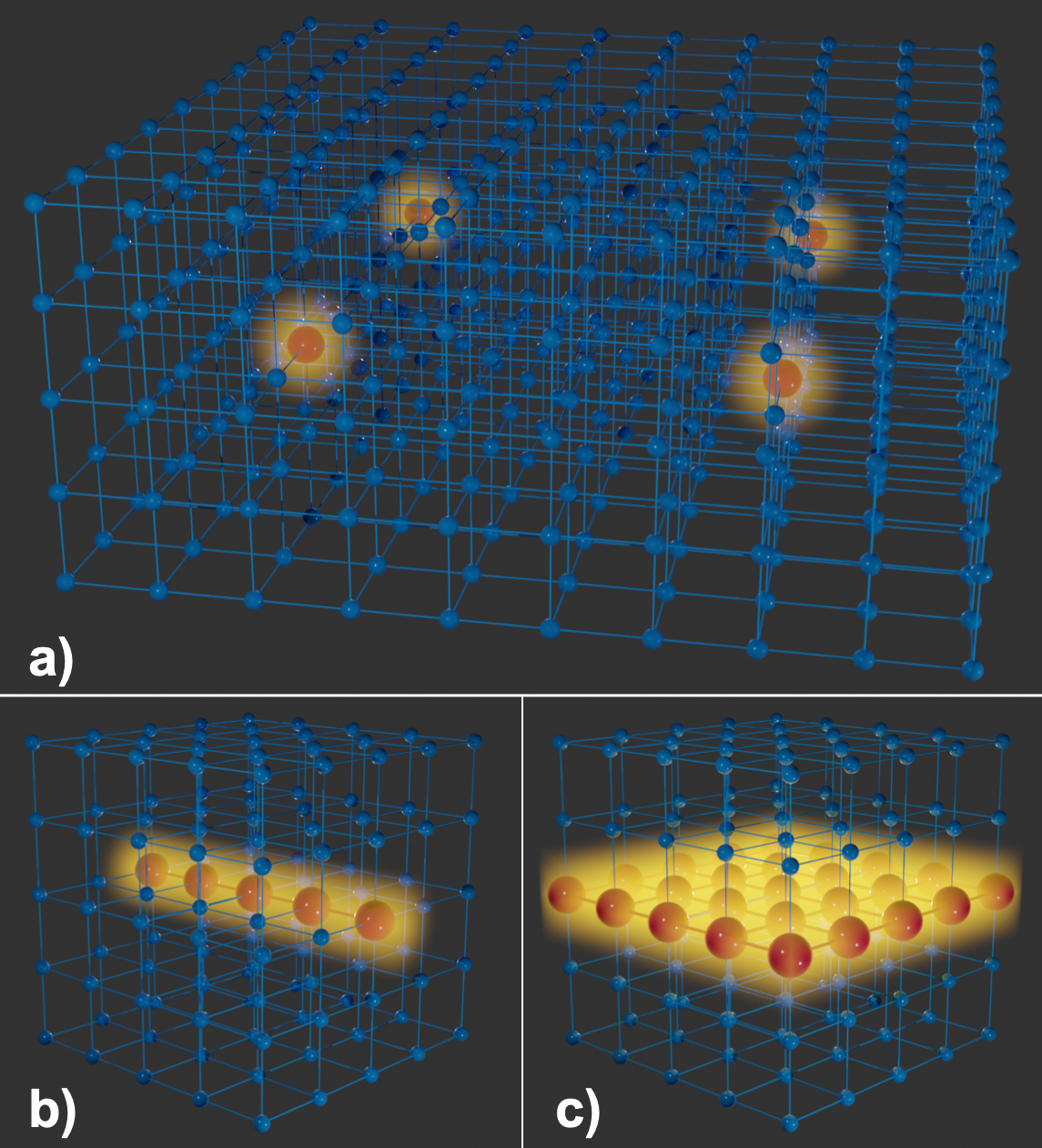}}
         \caption{Illustration of supercells of size $N=5$ in a $D=3$ system with various types of wave confinement dimensionalities $c$ induced by defect geometries of dimensionalities $d$. 
         Blue spheres correspond to regular unit cells, unit cells containing defects are red and confined waves are represented by yellow. 
         (a) Four supercells, each containing a point defect akin to a cavity \cite{Aspelmeyer2014Rev.Mod.Phys.}, trapping waves in all three dimensions, $d=0$, $c=3$.
         (b) Supercell with a linear defect, analogous to an optical waveguide or fiber, $d=1$, $c=2$.
         (c) Supercell with a planar defect, analogous to a 2D electron gas~\cite{Ando1982Rev.Mod.Phys.}, $d=2$, $c=1$.}
\label{fig:conf}
\end{figure}

Based on the geometric dimensionality $d$ of the defects, wave bands with different confinement dimensionality $c$ arise in the spectrum, as depicted in Fig.~\ref{fig:conf} for an $N=5$ supercell in $D=3$.
The confinement dimensionality $c$ of a wave determines the number of dimensions in which the wave is confined and mathematically corresponds to the codimension of the defect dimensionality: $c=D-d$.

We note that introducing defects in every unit cell of the supercell is equivalent to a new periodic structure with no defects at all ($d=3$) and supports only extended, \ie , unconfined ($c=0$), wave bands.

Fig.~\ref{fig:conf} represents highly idealized examples of simple defects. 
Real structures often contain multiple defects of various geometries oriented along different directions.
Such complex defect configurations allow for several different confinement dimensionalities $c$ and the problem of assigning the correct $c$ to a given band thus becomes highly nontrivial.
In order to understand the confinement in these real structures, a systematic method of identification and classification of confined bands is needed.

To analyze the confinement of waves, one must first determine the physical quantity corresponding to the notion of wave confinement for the given type of physical wave.
Without loss of generality, one can make this a real, non-negative quantity. 
For example, in electromagnetic systems this quantity could be the energy density, in electronic systems it could be the charge density, and similarly for other waves. 
To keep the description of our method general, in the following we will assume that this quantity has been identified and that it can be calculated for every band of interest.
We will denote this spatially dependent confinement quantity by $W(\vb x)$ and it will serve as the starting point of our analysis.

For scientifically interesting structures of moderate supercell sizes, the number of bands to be analyzed often reaches several hundreds (see, \eg , \cite{Woldering2014Phys.Rev.B, Hack2019Phys.Rev.B}).
It is therefore clearly impractical, borderline unfeasible, to analyze the confinement by visually inspecting the spatial distribution of $W(\vb x)$ for each band. 
Such an approach is also prone to human error and, due to its qualitative nature, hard to automate.
There is thus a clear need for a more quantitative approach towards the wave confinement analysis.

We employ $W(\vb x)$ to define the two key quantities for our scaling analysis.
First is the mode volume
%
\begin{equation}
\Vmode \colonequals \frac{\int\limits_{\Vsup}W(\vb x)\dd{V}}{\max\limits_{\vb x\in\Vsup} \{W(\vb x)\}},
\label{mvdef}
\end{equation}
%
where the integration is over the supercell volume $\Vsup$. 
We note that in literature the definitions of the mode volume may differ for various applications~\cite{Woldering2014Phys.Rev.B, Painter1999Science, Sauvan2013Phys.Rev.Lett., Kristensen2015Phys.Rev.A, Muljarov2016Phys.Rev.B}.
This difference is usually manifested in modification of the integration volume in Eq.~\eq{mvdef}.
In our case of a superlattice, it seems physically obvious to choose the supercell volume $\Vsup$ as the integration domain.
Moreover, as we show below, irrespective of the physical meaning of the quantity $\Vmode$, integrating over $\Vsup$ is crucial for obtaining the correct scaling behavior in our theory.
As an alternative quantity to the mode volume, the participation ratio~\cite{El-Dardiry2012Phys.Rev.B, Arceri2020Phys.Rev.Lett.} can be used, which also involves integration over $\Vsup$.
For details on this alternative, see Supplemental material.

The mode volume intuitively corresponds to the \textit{volume in which the wave is confined}.
On the other hand, it is important to also consider the problem from the converse point of view: \textit{How much energy of the wave is stored in a certain volume}? 
In order to quantify this notion, we define the  confinement energy
%
\begin{equation}
\Ecav \colonequals \int\limits_{\Vcav}W(\vb x)\dd{V},
\label{ecavdef}
\end{equation}
%
with the integration volume $\Vcav$ taken as the volume of one unit cell, centered around the defect.
An analogous quantity \bibnote{Termed SAW-likeness coefficient.} has been introduced by Ref.~\cite{Nardi2009Phys.Rev.B} to analyze confinement of surface acoustic waves.

For practical purposes, we employ normalized quantities, in the following:
%
\begin{equation}
\VV \colonequals \frac{\Vmode}{\Vsup}, \qquad \EE \colonequals \frac{\Ecav}{\Esup}.
\label{venorm}
\end{equation}
%
Here, $\Esup$ denotes the total amount of energy in the supercell.

It is clear that for a wave band confined within a cavity we expect simultaneously low $\VV$ and high $\EE $, while the opposite (high $\VV$, low $\EE$) is expected for a fully extended band.  
Nevertheless, there are no natural thresholds on how low $\VV$ and how high $\EE$ should be for a band to be clearly identified as confined. 
To overcome this ambiguity, we aim to analyze the behaviour of $\VV$ and $\EE$ with respect to the variation of the supercell size $N$, instead of their values themselves.
This technique is known as finite-size scaling~\cite{Abrahams1979Phys.Rev.Lett., Sheng2006book, Slevin2001Phys.Rev.Lett.}.

\begin{figure*}
	\includegraphics[width=\linewidth]{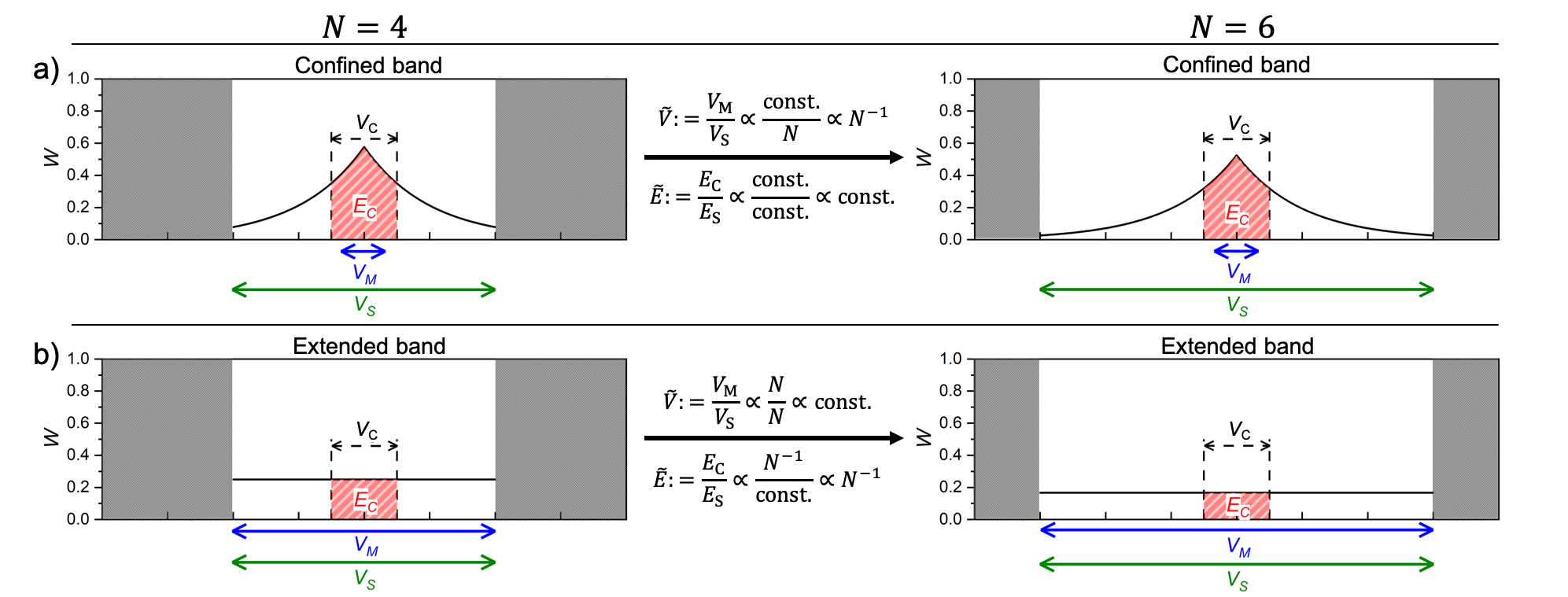}
	\caption{
	Illustration of our scaling argument in $D=1$. 
	An $N=4$ supercell is transformed into an $N=6$ supercell by adding unit cells to its boundary. 
	We investigate how the characteristic quantities $\widetilde V$ and $\widetilde E$ evolve upon this transition and obtain scaling relations for them. 
	(a) Confined band. 
	(b) Extended band.
	}
	\label{fig:scaling1D}
\end{figure*}

Our scaling argument is illustrated with the following didactic 1D model in Fig.~\ref{fig:scaling1D}. 
Let us consider a supercell with $N=4$ with a cavity in its center.
For a specific band with certain $\widetilde V$ and $\widetilde E$, we are interested in how adding more unit cells to the supercell boundary changes these values, and how these changes differ for a confined band as opposed to an extended band.
Therefore, we increase the supercell size to $N=6$. 
While scaling, we keep the total amount of energy within the supercell constant, \ie , $\Esup=\const$.
%
%

A confined band is depicted in Fig.~\ref{fig:scaling1D}(a) with its energy density decaying away from the cavity in the supercell center.
The size of the cavity to which the wave is confined and thus also the mode volume $\Vmode$ remain constant while scaling. 
However, the supercell volume $\Vsup$ has increased proportionally to $N$. 
Hence, the normalized quantity $\widetilde V$ decreases for the confined band as $N^{-1}$. 
Moreover, since the wave is confined within the cavity, it only 'feels' that additional unit cells have been added through its decaying tail.
This response clearly approaches zero as $N\ra\infty$ and, thus, in this limit $\widetilde E=\const$. 

An extended band is illustrated by a constant energy density distribution throughout the whole supercell in Fig.~\ref{fig:scaling1D}(b).
In this case, the behavior described above is the converse:
The volume occupied by the wave in the supercell grows proportionally to the supercell volume, resulting in the scaling $\widetilde V=\const$. 
The energy density extends homogeneously throughout the whole supercell and will further spread into the added unit cells as $N$ is increased. 
This leakage decreases the amount of confinement energy $\Ecav$ within $\Vcav$ as $N^{-1}$.

It is straightforward to extend the above scaling analysis to a $D$-dimensional system.
A band with a given confinement dimensionality $0\le c\le D$ obeys the following scaling relations, in the limit of $N\rightarrow\infty$:
\begin{equation}
\widetilde V=A N^{-c}, \qquad \widetilde E=B N^{c-D}.
\label{eq:vmodeecavD}
\end{equation}
Here, $A$ and $B$ are constants independent of $N$.
For derivation of Eqs.~\eqref{eq:vmodeecavD}, see Supplemental material.

One can directly calculate $\VV$ and $\EE$ from Eqs.~\eq{mvdef}-\eq{venorm} for each band, but because the constants $A, B$ are not known \textit{a priori}, Eqs.~\eq{eq:vmodeecavD} cannot be easily inverted to obtain $c$.
To accurately obtain $c$ from our scaling relations, we combine the two equations in~\eq{eq:vmodeecavD} into a ratio of the normalized mode volume raised to a judiciously chosen power $\alpha>0$ and the normalized confinement energy:
\begin{equation}
\frac{\widetilde V^\alpha}{\widetilde E} = C N^\kappa,
\label{eq:scaling}
\end{equation}
where the critical exponent $\kappa$ is given by
\begin{equation}
\kappa = -(\alpha + 1) c + D.
\label{eq:kappa}
\end{equation}

The equation \eq{eq:scaling} strictly holds only in the limit of very large $N$, with sub-leading order terms in $N$ present for finite supercell sizes.
However, in Eqs.~\eq{eq:scaling} and~\eq{eq:kappa}, we have introduced the auxiliary power $\alpha$, which can be utilized to effectively minimize the contribution of these sub-leading terms and thus identify the confinement of bands already for very small supercells.
Specifically, by suitably choosing the power $\alpha$, we can adjust the value of $\kappa$ so that it is negative for bands with certain $c$ and positive for the other bands. The value of $\kappa$, in turn, influences the scaling behavior of these bands as per Eq.~\eq{eq:scaling}. 

\begin{figure}
\includegraphics[width=\linewidth]{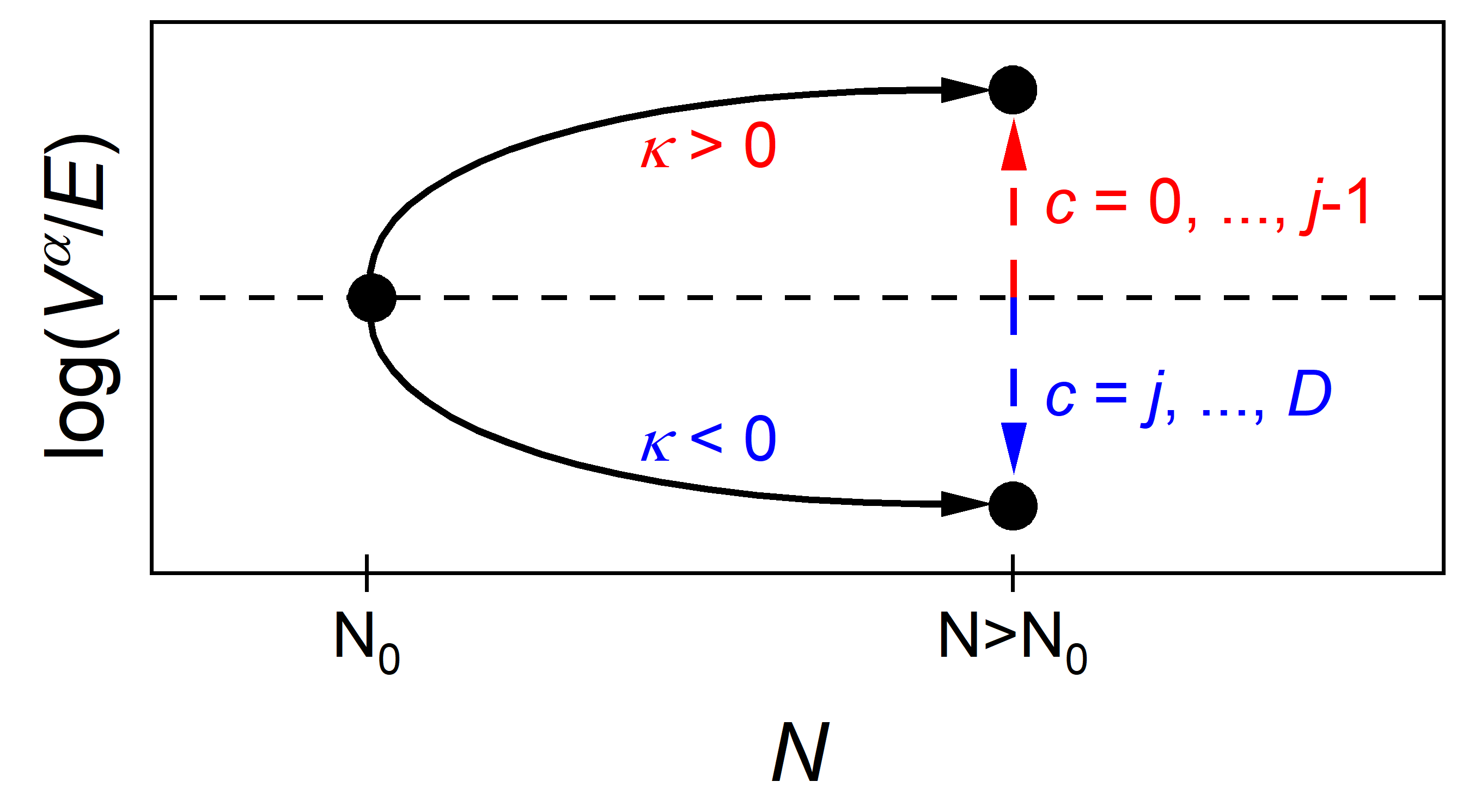}
         \caption{
         Flow diagram to identify bands confined in $c\ge j$ dimensions in a $D$-dimensional medium:
         The ratio $\VV^\alpha/\EE$ versus the system size $N$. 
         The auxiliary power $\alpha$ is chosen so that $\kappa<0$ for all $c\ge j$ and $\kappa>0$ for all $c<j$. 
         The sign of $\kappa$ determines if $\VV^\alpha/\EE$ increases or decreases with growing supercell size. 
         }
\label{fig:scaling_schematic}
\end{figure}

Our technique is illustrated in the flow diagram in Fig.~\ref{fig:scaling_schematic}. 
Investigating the confinement in a supercell of size $N$, for every $0< j \le D$, we have chosen $\alpha$ so that $\kappa<0$ for all $c\ge j$ and $\kappa>0$ for all $c< j$.
We are now able to distinguish between the bands with $c< j$ and $c\ge j$ by simply tracking the behavior of $\widetilde V^\alpha/\widetilde E$ as the supercell grows from a smaller reference size $N_0$ to the investigated size $N$.
The bands with negative $\kappa$ will move downwards in the graph, while the bands with positive $\kappa$ will shift upwards, in accordance with the Eq.~\eq{eq:scaling}.
By varying $j$ over all the integer values $0< j \le D$, this approach allows us to completely classify all wave bands in the spectrum based on their confinement dimensionality.

The unique values of $\alpha$ to maximize the accuracy of our technique for $D=1,2,3$ are listed in Table~\ref{table:kappa}.
For details on their calculation, see the Supplemental material.
%
%
\begin{table}[ht!]
\centering
\renewcommand\arraystretch{1.1}
 \begin{tabular}{ || c | c | c | c | c || }
 \cline{3-5}
   \multicolumn{2}{c|}{}& \multicolumn{3}{c||}{$D$}  \\ 
   \cline{3-5}
   \multicolumn{2}{c|}{} & 1 & 2  & 3     \\ 
   \hline
  \multirow{3}{*}{{\normalsize $j$\,}} 
  &  3  &   {  -- }    &  {  --    }      & {  \nicefrac{1}{5} }    \\ \cline{2-5}
  &   2 &   {  -- }    &  {  \nicefrac{1}{3} }      & {  1 }    \\ \cline{2-5}
  &   1 &   {  1}  &  {  3 }      & {  5}    \\ \hline
\end{tabular}
\caption{Auxiliary powers $\alpha$ to classify the confinement in systems of various dimension $D$. The value of $j<D$ indicates the confinement dimensionality being identified, as per Fig.~\ref{fig:scaling_schematic}.
}
 \label{table:kappa}
\end{table}

In case of the 1D example from Fig.~\ref{fig:scaling1D}, if we choose $\alpha=1$, corresponding to $D=1$ and $j=1$ from Table~\ref{table:kappa}, upon plotting $\tilde V/\tilde E$ versus the supercell size $N$ we will observe that the $c=1$ confined band moves down, while the $c=0$ extended band moves up in the flow diagram, analogously to Fig.~\ref{fig:scaling_schematic}. 

We now demonstrate our method on a 2D hexagonal BN with a nitrogen vacancy representing a point-like defect~\cite{Huang2012Phys.Rev.B}. 
Specifically, we investigate electronic confinement in a supercell of size $N=5$.
The band structure and charge densities of this quantum system were calculated using density functional theory~\cite{Kratzer2019Front.Chem.} implemented in the VASP code v6.1.1~\cite{Kresse1996ComputationalMaterialsScience}. 
For details see Supplemental material. 

\begin{figure}
\includegraphics[width=\linewidth]{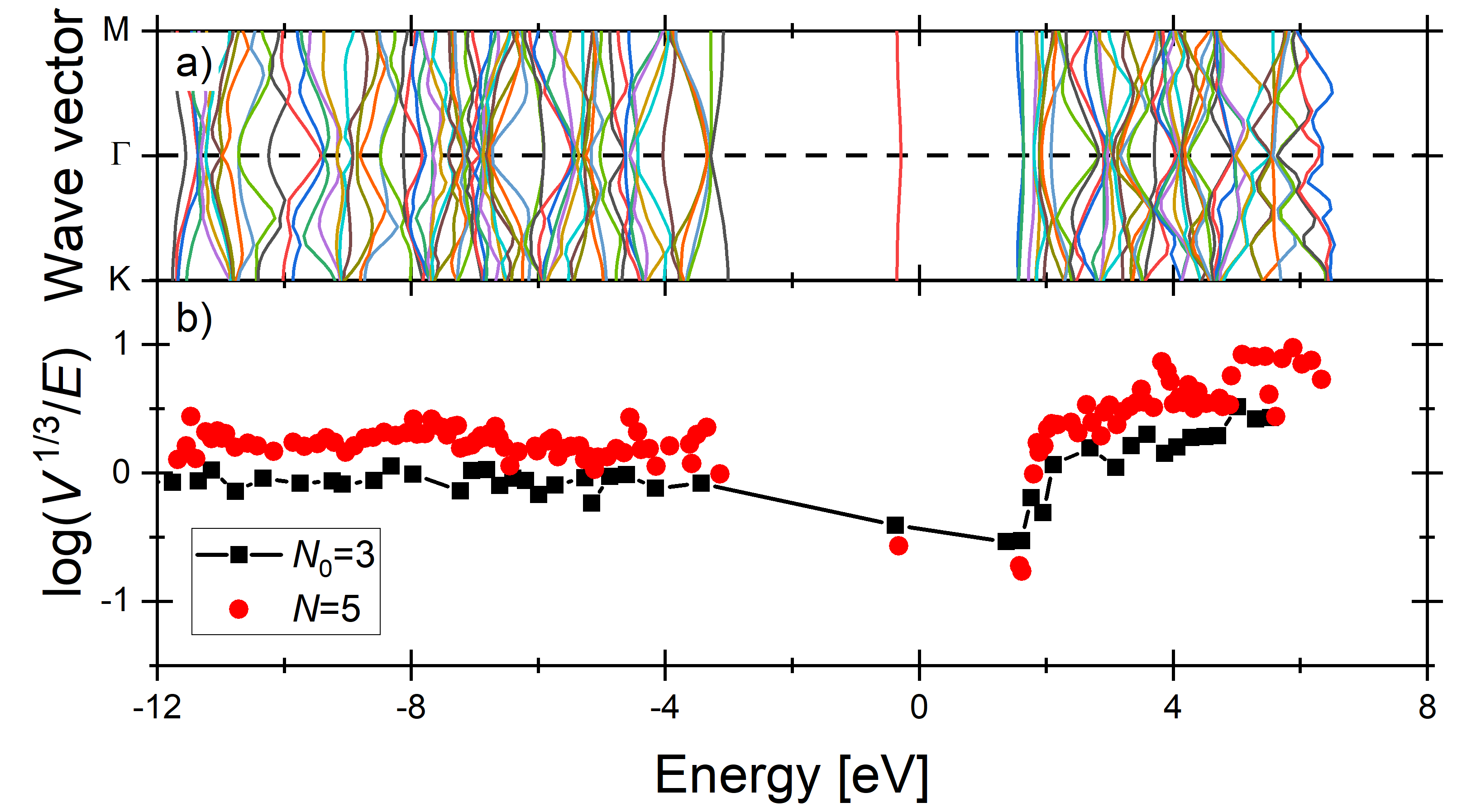}
         \caption{Our confinement analysis applied to a 2D quantum system: electronic waves in a hexagonal BN with nitrogen deficiency. 
         Zero energy is set at the Fermi level. 
         (a) Band structure of the $N=5$ supercell, flipped to have an energy abscissa.
         Bands are distinguished by colors for clarity. 
         (b) Scaling analysis of confinement: every band is represented by a point. 
         As the supercell size increases from $N_0=3$ to $N=5$, extended bands move upwards in the plot, and the three point-confined bands closest to zero energy shift down.
         }
\label{fig:DFTconfinement}
\end{figure}

The band structure of the system is shown in Fig.~\ref{fig:DFTconfinement}(a). 
Since the defect in our $D=2$ system has a point geometry, we only expect two different confinement dimensionalities to appear: the point-confined ($c=2$) and the extended ($c=0$) bands.
To distinguish between these, we choose $\alpha=\nicefrac{1}{3}$ from Table~\ref{table:kappa} and use a reference supercell of size $N_0=3$.
The plot of $\log(\VV^{\nicefrac{1}{3}}/\EE)$ for each band in Fig.~\ref{fig:DFTconfinement}(b) clearly shows that the majority of bands move upwards, identifying them as extended ($c=0$) bands according to our framework.
Additionally, three bands near zero energy move downwards and thus correspond to point-confined ($c=2$) bands. 

In our second example, we demonstrate our technique on light confinement in an $N=4$ supercell of a 3D inverse woodpile photonic crystal with two proximate line defects~\cite{Ho1994SolidStateCommun.,Woldering2014Phys.Rev.B, Devashish2019Phys.Rev.B, Hack2019Phys.Rev.B}.
The crossing point of these line defects represents a point defect.
We specifically study the acceptor-like structure investigated by Ref.~\cite{Woldering2014Phys.Rev.B}.
It is clear that the defects in this case have more complicated geometry than in the previous example: The system sustains point-confined ($c=3$) bands, line-confined ($c=2$) bands and extended ($c=0$) bands. 
The band structure and the energy densities were calculated with the MPB code~\cite{Johnson2001Opt.Express}.
For details on the structure and analysis, see Supplemental material. 

Fig.~\ref{fig:photonicconfinement} depicts the results of our analysis.
\begin{figure}
\includegraphics[width=\linewidth]{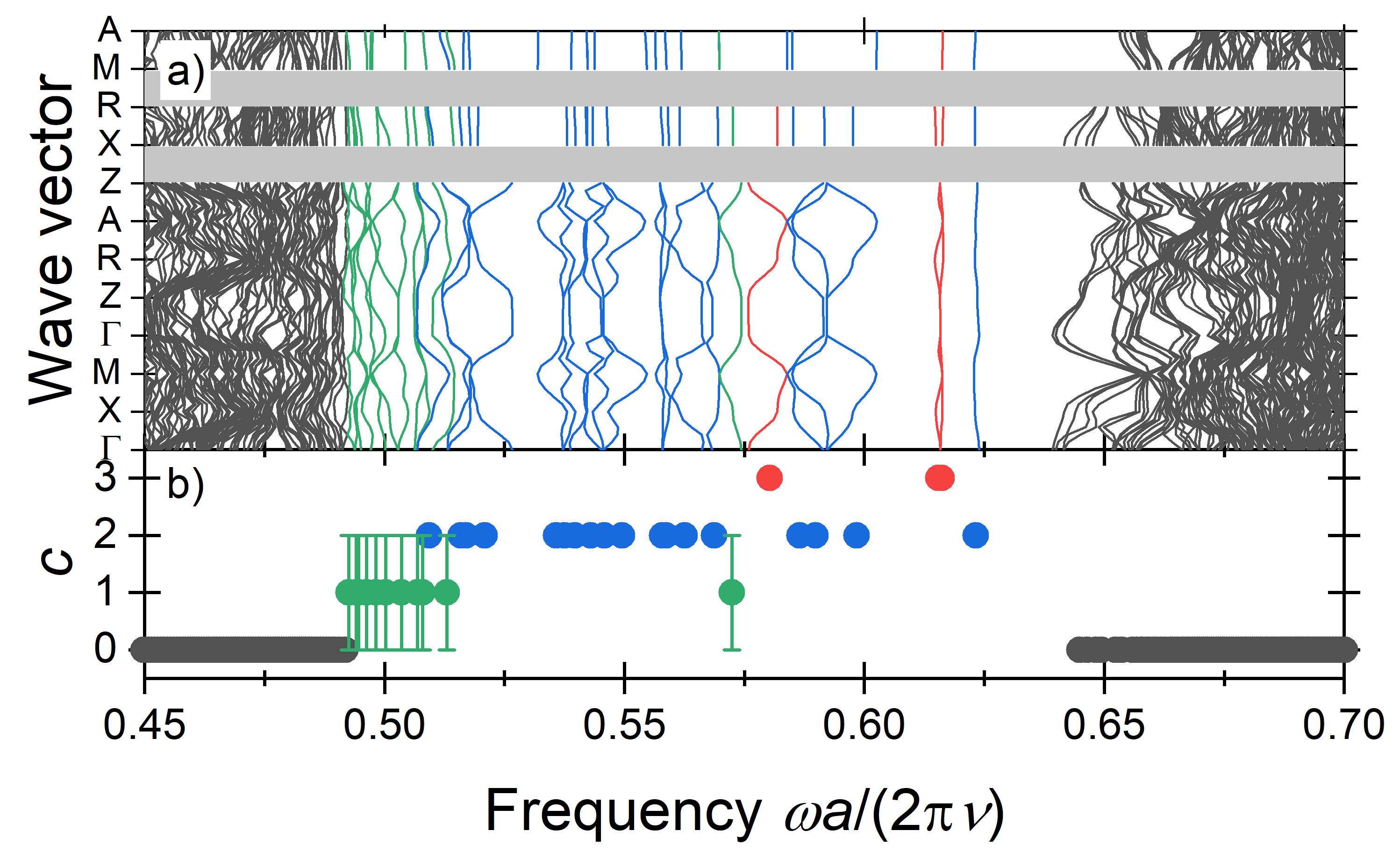}
         \caption{Scaling analysis of wave confinement in an $N=4$ supercell of a 3D inverse woodpile photonic crystal with two proximate line defects. 
         The horizontal axis displays the reduced frequency $\tilde\omega:=\omega a/(2\pi\nu)$, where $a$ is the lattice constant in the $y$ direction and $\nu$ denotes the speed of light.
         (a) Band structure. 
         (b) Confinement dimensionalities determined by our method for each band. 
         Error bars denote bands assigned to $c=1$, which should not occur in our structure but appears due to the small size of the studied supercell.
         }
\label{fig:photonicconfinement}
\end{figure} 
We unambiguously identify the confinement dimensionality $c$ of most bands. 
Additionally, several bands, mostly near $\tilde\omega\approx 0.5$, appear to be plane-confined ($c=1$). 
We know, however, that our defect geometry is linear and thus it does not support plane-confined bands.
We attribute this discrepancy to the small size of the studied supercell.
More specifically, these bands either have $c=0$, with the sub-leading order contributions to Eq.~\eq{eq:scaling} taking over for the studied supercell sizes, or they have $c=2$, with localization length in one of the dimensions still being larger than the reference supercell size $N_0$, thus appearing as only being confined in one dimension.
Nevertheless, our analysis is the first that distinguishes $c=2$ bands from those with $c=0$ in a 3D photonic superlattice. 

We also find three bands (red) with $c=3$.
Visual inspection confirms the point-confined character of the two bands at $\tilde\omega \approx 0.62$.
The inspection also shows that the band at $\tilde\omega \approx 0.56$ is confined in at least two dimensions, however it is visually not clear if the confinement is $c=2$ or $c=3$. 
Nevertheless, our analysis clearly disproves the statement in Ref.~\cite{Woldering2014Phys.Rev.B} that no point-confined bands exist in acceptor-like inverse woodpile crystals.

In this Letter we described a systematic scaling theory to analyze wave confinement in periodic media with functional defects, applicable to any type of physical waves. 
Already in one of the studied examples our technique uncovers physically new results, thus showcasing its power.
Our method is directly applicable to actively researched periodic structures and for optimization algorithms aiming to minimize or maximize specific types of wave confinement.

We thank Geert Brocks and Menno Bokdam for access to the CMS cluster with VASP code to perform DFT calculations. 
We also thank Geert Brocks for recommending hexagonal BN as a suitable material for illustrating our framework. 
Furthermore, we thank Sjoerd Hack for the introduction into the MPB software, and Lars Corbijn van Willenswaard and Manashee Adhikary for stimulating discussions on the physics of waves. 
This research is supported by the Shell-NWO/FOM programme "Computational Sciences for Energy Research" (CSER) and the MESA+ Institute for Nanotechnology, Applied Nanophotonics (ANP).

\bibliography{Cavity_superlattice_scaling} 

\begin{thebibliography}{59}%
\makeatletter
\providecommand \@ifxundefined [1]{%
 \@ifx{#1\undefined}
}%
\providecommand \@ifnum [1]{%
 \ifnum #1\expandafter \@firstoftwo
 \else \expandafter \@secondoftwo
 \fi
}%
\providecommand \@ifx [1]{%
 \ifx #1\expandafter \@firstoftwo
 \else \expandafter \@secondoftwo
 \fi
}%
\providecommand \natexlab [1]{#1}%
\providecommand \enquote  [1]{``#1''}%
\providecommand \bibnamefont  [1]{#1}%
\providecommand \bibfnamefont [1]{#1}%
\providecommand \citenamefont [1]{#1}%
\providecommand \href@noop [0]{\@secondoftwo}%
\providecommand \href [0]{\begingroup \@sanitize@url \@href}%
\providecommand \@href[1]{\@@startlink{#1}\@@href}%
\providecommand \@@href[1]{\endgroup#1\@@endlink}%
\providecommand \@sanitize@url [0]{\catcode `\\12\catcode `\$12\catcode
  `\&12\catcode `\#12\catcode `\^12\catcode `\_12\catcode `\%12\relax}%
\providecommand \@@startlink[1]{}%
\providecommand \@@endlink[0]{}%
\providecommand \url  [0]{\begingroup\@sanitize@url \@url }%
\providecommand \@url [1]{\endgroup\@href {#1}{\urlprefix }}%
\providecommand \urlprefix  [0]{URL }%
\providecommand \Eprint [0]{\href }%
\providecommand \doibase [0]{https://doi.org/}%
\providecommand \selectlanguage [0]{\@gobble}%
\providecommand \bibinfo  [0]{\@secondoftwo}%
\providecommand \bibfield  [0]{\@secondoftwo}%
\providecommand \translation [1]{[#1]}%
\providecommand \BibitemOpen [0]{}%
\providecommand \bibitemStop [0]{}%
\providecommand \bibitemNoStop [0]{.\EOS\space}%
\providecommand \EOS [0]{\spacefactor3000\relax}%
\providecommand \BibitemShut  [1]{\csname bibitem#1\endcsname}%
\let\auto@bib@innerbib\@empty
\bibitem [{\citenamefont {Marko{\v s}}\ and\ \citenamefont
  {Soukoulis}(2008)}]{Markos2008}%
  \BibitemOpen
  \bibfield  {author} {\bibinfo {author} {\bibfnamefont {P.}~\bibnamefont
  {Marko{\v s}}}\ and\ \bibinfo {author} {\bibfnamefont {C.~M.}\ \bibnamefont
  {Soukoulis}},\ }\href@noop {} {\emph {\bibinfo {title} {Wave Propagation:
  From Electrons to Photonic Crystals and Left-Handed Materials}}}\ (\bibinfo
  {publisher} {{Princeton University Press}},\ \bibinfo {address} {{Princeton ;
  Oxford}},\ \bibinfo {year} {2008})\BibitemShut {NoStop}%
\bibitem [{\citenamefont {Fink}\ \emph {et~al.}(2000)\citenamefont {Fink},
  \citenamefont {Cassereau}, \citenamefont {Derode}, \citenamefont {Prada},
  \citenamefont {Roux}, \citenamefont {Tanter}, \citenamefont {Thomas},\ and\
  \citenamefont {Wu}}]{Fink2000Rep.Prog.Phys.}%
  \BibitemOpen
  \bibfield  {author} {\bibinfo {author} {\bibfnamefont {M.}~\bibnamefont
  {Fink}}, \bibinfo {author} {\bibfnamefont {D.}~\bibnamefont {Cassereau}},
  \bibinfo {author} {\bibfnamefont {A.}~\bibnamefont {Derode}}, \bibinfo
  {author} {\bibfnamefont {C.}~\bibnamefont {Prada}}, \bibinfo {author}
  {\bibfnamefont {P.}~\bibnamefont {Roux}}, \bibinfo {author} {\bibfnamefont
  {M.}~\bibnamefont {Tanter}}, \bibinfo {author} {\bibfnamefont {J.-L.}\
  \bibnamefont {Thomas}},\ and\ \bibinfo {author} {\bibfnamefont
  {F.}~\bibnamefont {Wu}},\ }\href
  {https://doi.org/10.1088/0034-4885/63/12/202} {\bibfield  {journal} {\bibinfo
   {journal} {Rep. Prog. Phys.}\ }\textbf {\bibinfo {volume} {63}},\ \bibinfo
  {pages} {1933} (\bibinfo {year} {2000})}\BibitemShut {NoStop}%
\bibitem [{\citenamefont {Liu}(2000)}]{Liu2000Science}%
  \BibitemOpen
  \bibfield  {author} {\bibinfo {author} {\bibfnamefont {Z.}~\bibnamefont
  {Liu}},\ }\href {https://doi.org/10.1126/science.289.5485.1734} {\bibfield
  {journal} {\bibinfo  {journal} {Science}\ }\textbf {\bibinfo {volume}
  {289}},\ \bibinfo {pages} {1734} (\bibinfo {year} {2000})}\BibitemShut
  {NoStop}%
\bibitem [{\citenamefont {Maldovan}(2013)}]{Maldovan2013Nature}%
  \BibitemOpen
  \bibfield  {author} {\bibinfo {author} {\bibfnamefont {M.}~\bibnamefont
  {Maldovan}},\ }\href {https://doi.org/10.1038/nature12608} {\bibfield
  {journal} {\bibinfo  {journal} {Nature}\ }\textbf {\bibinfo {volume} {503}},\
  \bibinfo {pages} {209} (\bibinfo {year} {2013})}\BibitemShut {NoStop}%
\bibitem [{\citenamefont {Cummer}\ \emph {et~al.}(2016)\citenamefont {Cummer},
  \citenamefont {Christensen},\ and\ \citenamefont
  {Al{\`u}}}]{Cummer2016Nat.Rev.Mater.}%
  \BibitemOpen
  \bibfield  {author} {\bibinfo {author} {\bibfnamefont {S.~A.}\ \bibnamefont
  {Cummer}}, \bibinfo {author} {\bibfnamefont {J.}~\bibnamefont
  {Christensen}},\ and\ \bibinfo {author} {\bibfnamefont {A.}~\bibnamefont
  {Al{\`u}}},\ }\href {https://doi.org/10.1038/natrevmats.2016.1} {\bibfield
  {journal} {\bibinfo  {journal} {Nat. Rev. Mater.}\ }\textbf {\bibinfo
  {volume} {1}},\ \bibinfo {pages} {16001} (\bibinfo {year}
  {2016})}\BibitemShut {NoStop}%
\bibitem [{\citenamefont {Kruglyak}\ \emph {et~al.}(2010)\citenamefont
  {Kruglyak}, \citenamefont {Demokritov},\ and\ \citenamefont
  {Grundler}}]{Kruglyak2010J.Phys.D}%
  \BibitemOpen
  \bibfield  {author} {\bibinfo {author} {\bibfnamefont {V.~V.}\ \bibnamefont
  {Kruglyak}}, \bibinfo {author} {\bibfnamefont {S.~O.}\ \bibnamefont
  {Demokritov}},\ and\ \bibinfo {author} {\bibfnamefont {D.}~\bibnamefont
  {Grundler}},\ }\href {https://doi.org/10.1088/0022-3727/43/26/264001}
  {\bibfield  {journal} {\bibinfo  {journal} {J. Phys. D}\ }\textbf {\bibinfo
  {volume} {43}},\ \bibinfo {pages} {264001} (\bibinfo {year}
  {2010})}\BibitemShut {NoStop}%
\bibitem [{\citenamefont {Wagner}\ \emph {et~al.}(2016)\citenamefont {Wagner},
  \citenamefont {K{\'a}kay}, \citenamefont {Schultheiss}, \citenamefont
  {Henschke}, \citenamefont {Sebastian},\ and\ \citenamefont
  {Schultheiss}}]{Wagner2016Nat.Nanotechnol.}%
  \BibitemOpen
  \bibfield  {author} {\bibinfo {author} {\bibfnamefont {K.}~\bibnamefont
  {Wagner}}, \bibinfo {author} {\bibfnamefont {A.}~\bibnamefont {K{\'a}kay}},
  \bibinfo {author} {\bibfnamefont {K.}~\bibnamefont {Schultheiss}}, \bibinfo
  {author} {\bibfnamefont {A.}~\bibnamefont {Henschke}}, \bibinfo {author}
  {\bibfnamefont {T.}~\bibnamefont {Sebastian}},\ and\ \bibinfo {author}
  {\bibfnamefont {H.}~\bibnamefont {Schultheiss}},\ }\href
  {https://doi.org/10.1038/nnano.2015.339} {\bibfield  {journal} {\bibinfo
  {journal} {Nat. Nanotechnol.}\ }\textbf {\bibinfo {volume} {11}},\ \bibinfo
  {pages} {432} (\bibinfo {year} {2016})}\BibitemShut {NoStop}%
\bibitem [{\citenamefont {Klyukin}\ \emph {et~al.}(2018)\citenamefont
  {Klyukin}, \citenamefont {Tao}, \citenamefont {Tsymbal},\ and\ \citenamefont
  {Alexandrov}}]{Klyukin2018Phys.Rev.Lett.}%
  \BibitemOpen
  \bibfield  {author} {\bibinfo {author} {\bibfnamefont {K.}~\bibnamefont
  {Klyukin}}, \bibinfo {author} {\bibfnamefont {L.~L.}\ \bibnamefont {Tao}},
  \bibinfo {author} {\bibfnamefont {E.~Y.}\ \bibnamefont {Tsymbal}},\ and\
  \bibinfo {author} {\bibfnamefont {V.}~\bibnamefont {Alexandrov}},\ }\href
  {https://doi.org/10.1103/PhysRevLett.121.056601} {\bibfield  {journal}
  {\bibinfo  {journal} {Phys. Rev. Lett.}\ }\textbf {\bibinfo {volume} {121}},\
  \bibinfo {pages} {056601} (\bibinfo {year} {2018})}\BibitemShut {NoStop}%
\bibitem [{\citenamefont {Callahan}\ \emph {et~al.}(2013)\citenamefont
  {Callahan}, \citenamefont {Horowitz},\ and\ \citenamefont
  {Atwater}}]{Callahan2013Opt.Express}%
  \BibitemOpen
  \bibfield  {author} {\bibinfo {author} {\bibfnamefont {D.~M.}\ \bibnamefont
  {Callahan}}, \bibinfo {author} {\bibfnamefont {K.~A.~W.}\ \bibnamefont
  {Horowitz}},\ and\ \bibinfo {author} {\bibfnamefont {H.~A.}\ \bibnamefont
  {Atwater}},\ }\href {https://doi.org/10.1364/OE.21.030315} {\bibfield
  {journal} {\bibinfo  {journal} {Opt. Express}\ }\textbf {\bibinfo {volume}
  {21}},\ \bibinfo {pages} {30315} (\bibinfo {year} {2013})}\BibitemShut
  {NoStop}%
\bibitem [{\citenamefont {Tandaechanurat}\ \emph {et~al.}(2011)\citenamefont
  {Tandaechanurat}, \citenamefont {Ishida}, \citenamefont {Guimard},
  \citenamefont {Nomura}, \citenamefont {Iwamoto},\ and\ \citenamefont
  {Arakawa}}]{Tandaechanurat2011Nat.Photonics}%
  \BibitemOpen
  \bibfield  {author} {\bibinfo {author} {\bibfnamefont {A.}~\bibnamefont
  {Tandaechanurat}}, \bibinfo {author} {\bibfnamefont {S.}~\bibnamefont
  {Ishida}}, \bibinfo {author} {\bibfnamefont {D.}~\bibnamefont {Guimard}},
  \bibinfo {author} {\bibfnamefont {M.}~\bibnamefont {Nomura}}, \bibinfo
  {author} {\bibfnamefont {S.}~\bibnamefont {Iwamoto}},\ and\ \bibinfo {author}
  {\bibfnamefont {Y.}~\bibnamefont {Arakawa}},\ }\href
  {https://doi.org/10.1038/nphoton.2010.286} {\bibfield  {journal} {\bibinfo
  {journal} {Nat. Photonics}\ }\textbf {\bibinfo {volume} {5}},\ \bibinfo
  {pages} {91} (\bibinfo {year} {2011})}\BibitemShut {NoStop}%
\bibitem [{\citenamefont {Aspelmeyer}\ \emph {et~al.}(2014)\citenamefont
  {Aspelmeyer}, \citenamefont {Kippenberg},\ and\ \citenamefont
  {Marquardt}}]{Aspelmeyer2014Rev.Mod.Phys.}%
  \BibitemOpen
  \bibfield  {author} {\bibinfo {author} {\bibfnamefont {M.}~\bibnamefont
  {Aspelmeyer}}, \bibinfo {author} {\bibfnamefont {T.~J.}\ \bibnamefont
  {Kippenberg}},\ and\ \bibinfo {author} {\bibfnamefont {F.}~\bibnamefont
  {Marquardt}},\ }\href {https://doi.org/10.1103/RevModPhys.86.1391} {\bibfield
   {journal} {\bibinfo  {journal} {Rev. Mod. Phys.}\ }\textbf {\bibinfo
  {volume} {86}},\ \bibinfo {pages} {1391} (\bibinfo {year}
  {2014})}\BibitemShut {NoStop}%
\bibitem [{\citenamefont {Koenderink}\ \emph {et~al.}(2015)\citenamefont
  {Koenderink}, \citenamefont {Alu},\ and\ \citenamefont
  {Polman}}]{Koenderink2015Science}%
  \BibitemOpen
  \bibfield  {author} {\bibinfo {author} {\bibfnamefont {A.~F.}\ \bibnamefont
  {Koenderink}}, \bibinfo {author} {\bibfnamefont {A.}~\bibnamefont {Alu}},\
  and\ \bibinfo {author} {\bibfnamefont {A.}~\bibnamefont {Polman}},\ }\href
  {https://doi.org/10.1126/science.1261243} {\bibfield  {journal} {\bibinfo
  {journal} {Science}\ }\textbf {\bibinfo {volume} {348}},\ \bibinfo {pages}
  {516} (\bibinfo {year} {2015})}\BibitemShut {NoStop}%
\bibitem [{\citenamefont {Li}\ and\ \citenamefont
  {Fan}(2018)}]{Li2018Opt.Express}%
  \BibitemOpen
  \bibfield  {author} {\bibinfo {author} {\bibfnamefont {W.}~\bibnamefont
  {Li}}\ and\ \bibinfo {author} {\bibfnamefont {S.}~\bibnamefont {Fan}},\
  }\href {https://doi.org/10.1364/OE.26.015995} {\bibfield  {journal} {\bibinfo
   {journal} {Opt. Express}\ }\textbf {\bibinfo {volume} {26}},\ \bibinfo
  {pages} {15995} (\bibinfo {year} {2018})}\BibitemShut {NoStop}%
\bibitem [{\citenamefont {Wang}\ \emph {et~al.}(2020)\citenamefont {Wang},
  \citenamefont {Sciarrino}, \citenamefont {Laing},\ and\ \citenamefont
  {Thompson}}]{Wang2020Nat.Photonics}%
  \BibitemOpen
  \bibfield  {author} {\bibinfo {author} {\bibfnamefont {J.}~\bibnamefont
  {Wang}}, \bibinfo {author} {\bibfnamefont {F.}~\bibnamefont {Sciarrino}},
  \bibinfo {author} {\bibfnamefont {A.}~\bibnamefont {Laing}},\ and\ \bibinfo
  {author} {\bibfnamefont {M.~G.}\ \bibnamefont {Thompson}},\ }\href
  {https://doi.org/10.1038/s41566-019-0532-1} {\bibfield  {journal} {\bibinfo
  {journal} {Nat. Photonics}\ }\textbf {\bibinfo {volume} {14}},\ \bibinfo
  {pages} {273} (\bibinfo {year} {2020})}\BibitemShut {NoStop}%
\bibitem [{\citenamefont {Uppu}\ \emph {et~al.}(2021)\citenamefont {Uppu},
  \citenamefont {Adhikary}, \citenamefont {Harteveld},\ and\ \citenamefont
  {Vos}}]{Uppu2021Phys.Rev.Lett.}%
  \BibitemOpen
  \bibfield  {author} {\bibinfo {author} {\bibfnamefont {R.}~\bibnamefont
  {Uppu}}, \bibinfo {author} {\bibfnamefont {M.}~\bibnamefont {Adhikary}},
  \bibinfo {author} {\bibfnamefont {C.~A.~M.}\ \bibnamefont {Harteveld}},\ and\
  \bibinfo {author} {\bibfnamefont {W.~L.}\ \bibnamefont {Vos}},\ }\href
  {https://doi.org/10.1103/PhysRevLett.126.177402} {\bibfield  {journal}
  {\bibinfo  {journal} {Phys. Rev. Lett.}\ }\textbf {\bibinfo {volume} {126}},\
  \bibinfo {pages} {177402} (\bibinfo {year} {2021})}\BibitemShut {NoStop}%
\bibitem [{\citenamefont {Arceri}\ and\ \citenamefont
  {Corwin}(2020)}]{Arceri2020Phys.Rev.Lett.}%
  \BibitemOpen
  \bibfield  {author} {\bibinfo {author} {\bibfnamefont {F.}~\bibnamefont
  {Arceri}}\ and\ \bibinfo {author} {\bibfnamefont {E.~I.}\ \bibnamefont
  {Corwin}},\ }\href {https://doi.org/10.1103/PhysRevLett.124.238002}
  {\bibfield  {journal} {\bibinfo  {journal} {Phys. Rev. Lett.}\ }\textbf
  {\bibinfo {volume} {124}},\ \bibinfo {pages} {238002} (\bibinfo {year}
  {2020})}\BibitemShut {NoStop}%
\bibitem [{\citenamefont {Busch}\ \emph {et~al.}(2007)\citenamefont {Busch},
  \citenamefont {{von Freymann}}, \citenamefont {Linden}, \citenamefont
  {Mingaleev}, \citenamefont {Tkeshelashvili},\ and\ \citenamefont
  {Wegener}}]{Busch2007Phys.Rep.}%
  \BibitemOpen
  \bibfield  {author} {\bibinfo {author} {\bibfnamefont {K.}~\bibnamefont
  {Busch}}, \bibinfo {author} {\bibfnamefont {G.}~\bibnamefont {{von
  Freymann}}}, \bibinfo {author} {\bibfnamefont {S.}~\bibnamefont {Linden}},
  \bibinfo {author} {\bibfnamefont {S.}~\bibnamefont {Mingaleev}}, \bibinfo
  {author} {\bibfnamefont {L.}~\bibnamefont {Tkeshelashvili}},\ and\ \bibinfo
  {author} {\bibfnamefont {M.}~\bibnamefont {Wegener}},\ }\href
  {https://doi.org/10.1016/j.physrep.2007.02.011} {\bibfield  {journal}
  {\bibinfo  {journal} {Phys. Rep.}\ }\textbf {\bibinfo {volume} {444}},\
  \bibinfo {pages} {101} (\bibinfo {year} {2007})}\BibitemShut {NoStop}%
\bibitem [{\citenamefont {Woldering}\ \emph {et~al.}(2014)\citenamefont
  {Woldering}, \citenamefont {Mosk},\ and\ \citenamefont
  {Vos}}]{Woldering2014Phys.Rev.B}%
  \BibitemOpen
  \bibfield  {author} {\bibinfo {author} {\bibfnamefont {L.~A.}\ \bibnamefont
  {Woldering}}, \bibinfo {author} {\bibfnamefont {A.~P.}\ \bibnamefont
  {Mosk}},\ and\ \bibinfo {author} {\bibfnamefont {W.~L.}\ \bibnamefont
  {Vos}},\ }\href {https://doi.org/10.1103/PhysRevB.90.115140} {\bibfield
  {journal} {\bibinfo  {journal} {Phys. Rev. B}\ }\textbf {\bibinfo {volume}
  {90}},\ \bibinfo {pages} {115140} (\bibinfo {year} {2014})}\BibitemShut
  {NoStop}%
\bibitem [{\citenamefont {Hack}\ \emph {et~al.}(2019)\citenamefont {Hack},
  \citenamefont {{van der Vegt}},\ and\ \citenamefont
  {Vos}}]{Hack2019Phys.Rev.B}%
  \BibitemOpen
  \bibfield  {author} {\bibinfo {author} {\bibfnamefont {S.~A.}\ \bibnamefont
  {Hack}}, \bibinfo {author} {\bibfnamefont {J.~J.~W.}\ \bibnamefont {{van der
  Vegt}}},\ and\ \bibinfo {author} {\bibfnamefont {W.~L.}\ \bibnamefont
  {Vos}},\ }\href {https://doi.org/10.1103/PhysRevB.99.115308} {\bibfield
  {journal} {\bibinfo  {journal} {Phys. Rev. B}\ }\textbf {\bibinfo {volume}
  {99}},\ \bibinfo {pages} {115308} (\bibinfo {year} {2019})}\BibitemShut
  {NoStop}%
\bibitem [{\citenamefont {Economou}(2010)}]{Economou2010}%
  \BibitemOpen
  \bibfield  {author} {\bibinfo {author} {\bibfnamefont {E.~N.}\ \bibnamefont
  {Economou}},\ }\href {https://doi.org/10.1007/978-3-642-02069-8} {\emph
  {\bibinfo {title} {The {{Physics}} of {{Solids}}}}},\ Graduate {{Texts}} in
  {{Physics}}\ (\bibinfo  {publisher} {{Springer Berlin Heidelberg}},\ \bibinfo
  {address} {{Berlin, Heidelberg}},\ \bibinfo {year} {2010})\BibitemShut
  {NoStop}%
\bibitem [{\citenamefont {Shao}(2008)}]{Shao2008J.Phys.Chem.C}%
  \BibitemOpen
  \bibfield  {author} {\bibinfo {author} {\bibfnamefont {G.}~\bibnamefont
  {Shao}},\ }\href {https://doi.org/10.1021/jp8043797} {\bibfield  {journal}
  {\bibinfo  {journal} {J. Phys. Chem. C}\ }\textbf {\bibinfo {volume} {112}},\
  \bibinfo {pages} {18677} (\bibinfo {year} {2008})}\BibitemShut {NoStop}%
\bibitem [{\citenamefont {Pashartis}\ and\ \citenamefont
  {Rubel}(2017{\natexlab{a}})}]{Pashartis2017Phys.Rev.Appl.}%
  \BibitemOpen
  \bibfield  {author} {\bibinfo {author} {\bibfnamefont {C.}~\bibnamefont
  {Pashartis}}\ and\ \bibinfo {author} {\bibfnamefont {O.}~\bibnamefont
  {Rubel}},\ }\href {https://doi.org/10.1103/PhysRevApplied.7.064011}
  {\bibfield  {journal} {\bibinfo  {journal} {Phys. Rev. Appl.}\ }\textbf
  {\bibinfo {volume} {7}},\ \bibinfo {pages} {064011} (\bibinfo {year}
  {2017}{\natexlab{a}})}\BibitemShut {NoStop}%
\bibitem [{\citenamefont {Pashartis}\ and\ \citenamefont
  {Rubel}(2017{\natexlab{b}})}]{Pashartis2017Phys.Rev.B}%
  \BibitemOpen
  \bibfield  {author} {\bibinfo {author} {\bibfnamefont {C.}~\bibnamefont
  {Pashartis}}\ and\ \bibinfo {author} {\bibfnamefont {O.}~\bibnamefont
  {Rubel}},\ }\href {https://doi.org/10.1103/PhysRevB.96.155209} {\bibfield
  {journal} {\bibinfo  {journal} {Phys. Rev. B}\ }\textbf {\bibinfo {volume}
  {96}},\ \bibinfo {pages} {155209} (\bibinfo {year}
  {2017}{\natexlab{b}})}\BibitemShut {NoStop}%
\bibitem [{\citenamefont {Zhang}\ \emph {et~al.}(2011)\citenamefont {Zhang},
  \citenamefont {Wang}, \citenamefont {Zhu},\ and\ \citenamefont
  {Dobrovitski}}]{Zhang2011Phys.Rev.B}%
  \BibitemOpen
  \bibfield  {author} {\bibinfo {author} {\bibfnamefont {J.}~\bibnamefont
  {Zhang}}, \bibinfo {author} {\bibfnamefont {C.-Z.}\ \bibnamefont {Wang}},
  \bibinfo {author} {\bibfnamefont {Z.~Z.}\ \bibnamefont {Zhu}},\ and\ \bibinfo
  {author} {\bibfnamefont {V.~V.}\ \bibnamefont {Dobrovitski}},\ }\href
  {https://doi.org/10.1103/PhysRevB.84.035211} {\bibfield  {journal} {\bibinfo
  {journal} {Phys. Rev. B}\ }\textbf {\bibinfo {volume} {84}},\ \bibinfo
  {pages} {035211} (\bibinfo {year} {2011})}\BibitemShut {NoStop}%
\bibitem [{\citenamefont {Demokritov}(2017)}]{Demokritov2017book}%
  \BibitemOpen
  \bibinfo {editor} {\bibfnamefont {S.~O.}\ \bibnamefont {Demokritov}},\ ed.,\
  \href@noop {} {\emph {\bibinfo {title} {Spin Wave Confinement: Propagating
  Waves}}},\ \bibinfo {edition} {second edition}\ ed.\ (\bibinfo  {publisher}
  {{Pan Stanford Publishing}},\ \bibinfo {address} {{Singapur}},\ \bibinfo
  {year} {2017})\BibitemShut {NoStop}%
\bibitem [{\citenamefont {Tartakovskaya}\ \emph {et~al.}(2016)\citenamefont
  {Tartakovskaya}, \citenamefont {{Pardavi-Horvath}},\ and\ \citenamefont
  {McMichael}}]{Tartakovskaya2016Phys.Rev.B}%
  \BibitemOpen
  \bibfield  {author} {\bibinfo {author} {\bibfnamefont {E.~V.}\ \bibnamefont
  {Tartakovskaya}}, \bibinfo {author} {\bibfnamefont {M.}~\bibnamefont
  {{Pardavi-Horvath}}},\ and\ \bibinfo {author} {\bibfnamefont {R.~D.}\
  \bibnamefont {McMichael}},\ }\href
  {https://doi.org/10.1103/PhysRevB.93.214436} {\bibfield  {journal} {\bibinfo
  {journal} {Phys. Rev. B}\ }\textbf {\bibinfo {volume} {93}},\ \bibinfo
  {pages} {214436} (\bibinfo {year} {2016})}\BibitemShut {NoStop}%
\bibitem [{\citenamefont {Krioukov}\ \emph {et~al.}(2002)\citenamefont
  {Krioukov}, \citenamefont {Klunder}, \citenamefont {Driessen}, \citenamefont
  {Greve},\ and\ \citenamefont {Otto}}]{Krioukov2002Opt.Lett.}%
  \BibitemOpen
  \bibfield  {author} {\bibinfo {author} {\bibfnamefont {E.}~\bibnamefont
  {Krioukov}}, \bibinfo {author} {\bibfnamefont {D.~J.~W.}\ \bibnamefont
  {Klunder}}, \bibinfo {author} {\bibfnamefont {A.}~\bibnamefont {Driessen}},
  \bibinfo {author} {\bibfnamefont {J.}~\bibnamefont {Greve}},\ and\ \bibinfo
  {author} {\bibfnamefont {C.}~\bibnamefont {Otto}},\ }\href
  {https://doi.org/10.1364/OL.27.000512} {\bibfield  {journal} {\bibinfo
  {journal} {Opt. Lett.}\ }\textbf {\bibinfo {volume} {27}},\ \bibinfo {pages}
  {512} (\bibinfo {year} {2002})}\BibitemShut {NoStop}%
\bibitem [{\citenamefont {Baba}(2008)}]{Baba2008Nat.Photonics}%
  \BibitemOpen
  \bibfield  {author} {\bibinfo {author} {\bibfnamefont {T.}~\bibnamefont
  {Baba}},\ }\href {https://doi.org/10.1038/nphoton.2008.146} {\bibfield
  {journal} {\bibinfo  {journal} {Nat. Photonics}\ }\textbf {\bibinfo {volume}
  {2}},\ \bibinfo {pages} {465} (\bibinfo {year} {2008})}\BibitemShut {NoStop}%
\bibitem [{\citenamefont {Noda}\ \emph {et~al.}(2000)\citenamefont {Noda},
  \citenamefont {Chutinan},\ and\ \citenamefont {Imada}}]{Noda2000Nature}%
  \BibitemOpen
  \bibfield  {author} {\bibinfo {author} {\bibfnamefont {S.}~\bibnamefont
  {Noda}}, \bibinfo {author} {\bibfnamefont {A.}~\bibnamefont {Chutinan}},\
  and\ \bibinfo {author} {\bibfnamefont {M.}~\bibnamefont {Imada}},\ }\href
  {https://doi.org/10.1038/35036532} {\bibfield  {journal} {\bibinfo  {journal}
  {Nature}\ }\textbf {\bibinfo {volume} {407}},\ \bibinfo {pages} {608}
  (\bibinfo {year} {2000})}\BibitemShut {NoStop}%
\bibitem [{\citenamefont {G{\'e}rard}\ \emph {et~al.}(1998)\citenamefont
  {G{\'e}rard}, \citenamefont {Sermage}, \citenamefont {Gayral}, \citenamefont
  {Legrand}, \citenamefont {Costard},\ and\ \citenamefont
  {{Thierry-Mieg}}}]{Gerard1998Phys.Rev.Lett.}%
  \BibitemOpen
  \bibfield  {author} {\bibinfo {author} {\bibfnamefont {J.~M.}\ \bibnamefont
  {G{\'e}rard}}, \bibinfo {author} {\bibfnamefont {B.}~\bibnamefont {Sermage}},
  \bibinfo {author} {\bibfnamefont {B.}~\bibnamefont {Gayral}}, \bibinfo
  {author} {\bibfnamefont {B.}~\bibnamefont {Legrand}}, \bibinfo {author}
  {\bibfnamefont {E.}~\bibnamefont {Costard}},\ and\ \bibinfo {author}
  {\bibfnamefont {V.}~\bibnamefont {{Thierry-Mieg}}},\ }\href
  {https://doi.org/10.1103/PhysRevLett.81.1110} {\bibfield  {journal} {\bibinfo
   {journal} {Phys. Rev. Lett.}\ }\textbf {\bibinfo {volume} {81}},\ \bibinfo
  {pages} {1110} (\bibinfo {year} {1998})}\BibitemShut {NoStop}%
\bibitem [{\citenamefont {Michler}(2003)}]{Michler2003}%
  \BibitemOpen
  \bibinfo {editor} {\bibfnamefont {P.}~\bibnamefont {Michler}},\ ed.,\
  \href@noop {} {\emph {\bibinfo {title} {Single Quantum Dots: Fundamentals,
  Applications, and New Concepts}}},\ \bibinfo {series} {Topics in Applied
  Physics}\ No.\ \bibinfo {number} {v. 90}\ (\bibinfo  {publisher}
  {{Springer-Verlag}},\ \bibinfo {address} {{Berlin, Heidelberg, New York}},\
  \bibinfo {year} {2003})\BibitemShut {NoStop}%
\bibitem [{\citenamefont {Reithmaier}\ \emph {et~al.}(2004)\citenamefont
  {Reithmaier}, \citenamefont {S{\k{e}}k}, \citenamefont {L{\"o}ffler},
  \citenamefont {Hofmann}, \citenamefont {Kuhn}, \citenamefont {Reitzenstein},
  \citenamefont {Keldysh}, \citenamefont {Kulakovskii}, \citenamefont
  {Reinecke},\ and\ \citenamefont {Forchel}}]{Reithmaier2004Nature}%
  \BibitemOpen
  \bibfield  {author} {\bibinfo {author} {\bibfnamefont {J.~P.}\ \bibnamefont
  {Reithmaier}}, \bibinfo {author} {\bibfnamefont {G.}~\bibnamefont
  {S{\k{e}}k}}, \bibinfo {author} {\bibfnamefont {A.}~\bibnamefont
  {L{\"o}ffler}}, \bibinfo {author} {\bibfnamefont {C.}~\bibnamefont
  {Hofmann}}, \bibinfo {author} {\bibfnamefont {S.}~\bibnamefont {Kuhn}},
  \bibinfo {author} {\bibfnamefont {S.}~\bibnamefont {Reitzenstein}}, \bibinfo
  {author} {\bibfnamefont {L.~V.}\ \bibnamefont {Keldysh}}, \bibinfo {author}
  {\bibfnamefont {V.~D.}\ \bibnamefont {Kulakovskii}}, \bibinfo {author}
  {\bibfnamefont {T.~L.}\ \bibnamefont {Reinecke}},\ and\ \bibinfo {author}
  {\bibfnamefont {A.}~\bibnamefont {Forchel}},\ }\href
  {https://doi.org/10.1038/nature02969} {\bibfield  {journal} {\bibinfo
  {journal} {Nature}\ }\textbf {\bibinfo {volume} {432}},\ \bibinfo {pages}
  {197} (\bibinfo {year} {2004})}\BibitemShut {NoStop}%
\bibitem [{\citenamefont {Yoshie}\ \emph {et~al.}(2004)\citenamefont {Yoshie},
  \citenamefont {Scherer}, \citenamefont {Hendrickson}, \citenamefont
  {Khitrova}, \citenamefont {Gibbs}, \citenamefont {Rupper}, \citenamefont
  {Ell}, \citenamefont {Shchekin},\ and\ \citenamefont
  {Deppe}}]{Yoshie2004Nature}%
  \BibitemOpen
  \bibfield  {author} {\bibinfo {author} {\bibfnamefont {T.}~\bibnamefont
  {Yoshie}}, \bibinfo {author} {\bibfnamefont {A.}~\bibnamefont {Scherer}},
  \bibinfo {author} {\bibfnamefont {J.}~\bibnamefont {Hendrickson}}, \bibinfo
  {author} {\bibfnamefont {G.}~\bibnamefont {Khitrova}}, \bibinfo {author}
  {\bibfnamefont {H.~M.}\ \bibnamefont {Gibbs}}, \bibinfo {author}
  {\bibfnamefont {G.}~\bibnamefont {Rupper}}, \bibinfo {author} {\bibfnamefont
  {C.}~\bibnamefont {Ell}}, \bibinfo {author} {\bibfnamefont {O.~B.}\
  \bibnamefont {Shchekin}},\ and\ \bibinfo {author} {\bibfnamefont {D.~G.}\
  \bibnamefont {Deppe}},\ }\href {https://doi.org/10.1038/nature03119}
  {\bibfield  {journal} {\bibinfo  {journal} {Nature}\ }\textbf {\bibinfo
  {volume} {432}},\ \bibinfo {pages} {200} (\bibinfo {year}
  {2004})}\BibitemShut {NoStop}%
\bibitem [{\citenamefont {Peter}\ \emph {et~al.}(2005)\citenamefont {Peter},
  \citenamefont {Senellart}, \citenamefont {Martrou}, \citenamefont
  {Lema{\^i}tre}, \citenamefont {Hours}, \citenamefont {G{\'e}rard},\ and\
  \citenamefont {Bloch}}]{Peter2005Phys.Rev.Lett.}%
  \BibitemOpen
  \bibfield  {author} {\bibinfo {author} {\bibfnamefont {E.}~\bibnamefont
  {Peter}}, \bibinfo {author} {\bibfnamefont {P.}~\bibnamefont {Senellart}},
  \bibinfo {author} {\bibfnamefont {D.}~\bibnamefont {Martrou}}, \bibinfo
  {author} {\bibfnamefont {A.}~\bibnamefont {Lema{\^i}tre}}, \bibinfo {author}
  {\bibfnamefont {J.}~\bibnamefont {Hours}}, \bibinfo {author} {\bibfnamefont
  {J.~M.}\ \bibnamefont {G{\'e}rard}},\ and\ \bibinfo {author} {\bibfnamefont
  {J.}~\bibnamefont {Bloch}},\ }\href
  {https://doi.org/10.1103/PhysRevLett.95.067401} {\bibfield  {journal}
  {\bibinfo  {journal} {Phys. Rev. Lett.}\ }\textbf {\bibinfo {volume} {95}},\
  \bibinfo {pages} {067401} (\bibinfo {year} {2005})}\BibitemShut {NoStop}%
\bibitem [{\citenamefont {Russell}\ \emph {et~al.}(2003)\citenamefont
  {Russell}, \citenamefont {Marin}, \citenamefont {Diez}, \citenamefont
  {Guenneau},\ and\ \citenamefont {Movchan}}]{Russell2003Opt.Express}%
  \BibitemOpen
  \bibfield  {author} {\bibinfo {author} {\bibfnamefont {P.}~\bibnamefont
  {Russell}}, \bibinfo {author} {\bibfnamefont {E.}~\bibnamefont {Marin}},
  \bibinfo {author} {\bibfnamefont {A.}~\bibnamefont {Diez}}, \bibinfo {author}
  {\bibfnamefont {S.}~\bibnamefont {Guenneau}},\ and\ \bibinfo {author}
  {\bibfnamefont {A.}~\bibnamefont {Movchan}},\ }\href
  {https://doi.org/10.1364/OE.11.002555} {\bibfield  {journal} {\bibinfo
  {journal} {Opt. Express}\ }\textbf {\bibinfo {volume} {11}},\ \bibinfo
  {pages} {2555} (\bibinfo {year} {2003})}\BibitemShut {NoStop}%
\bibitem [{\citenamefont {Painter}(1999)}]{Painter1999Science}%
  \BibitemOpen
  \bibfield  {author} {\bibinfo {author} {\bibfnamefont {O.}~\bibnamefont
  {Painter}},\ }\href {https://doi.org/10.1126/science.284.5421.1819}
  {\bibfield  {journal} {\bibinfo  {journal} {Science}\ }\textbf {\bibinfo
  {volume} {284}},\ \bibinfo {pages} {1819} (\bibinfo {year}
  {1999})}\BibitemShut {NoStop}%
\bibitem [{\citenamefont {Sauvan}\ \emph {et~al.}(2013)\citenamefont {Sauvan},
  \citenamefont {Hugonin}, \citenamefont {Maksymov},\ and\ \citenamefont
  {Lalanne}}]{Sauvan2013Phys.Rev.Lett.}%
  \BibitemOpen
  \bibfield  {author} {\bibinfo {author} {\bibfnamefont {C.}~\bibnamefont
  {Sauvan}}, \bibinfo {author} {\bibfnamefont {J.~P.}\ \bibnamefont {Hugonin}},
  \bibinfo {author} {\bibfnamefont {I.~S.}\ \bibnamefont {Maksymov}},\ and\
  \bibinfo {author} {\bibfnamefont {P.}~\bibnamefont {Lalanne}},\ }\href
  {https://doi.org/10.1103/PhysRevLett.110.237401} {\bibfield  {journal}
  {\bibinfo  {journal} {Phys. Rev. Lett.}\ }\textbf {\bibinfo {volume} {110}},\
  \bibinfo {pages} {237401} (\bibinfo {year} {2013})}\BibitemShut {NoStop}%
\bibitem [{\citenamefont {Kristensen}\ \emph {et~al.}(2015)\citenamefont
  {Kristensen}, \citenamefont {Ge},\ and\ \citenamefont
  {Hughes}}]{Kristensen2015Phys.Rev.A}%
  \BibitemOpen
  \bibfield  {author} {\bibinfo {author} {\bibfnamefont {P.~T.}\ \bibnamefont
  {Kristensen}}, \bibinfo {author} {\bibfnamefont {R.-C.}\ \bibnamefont {Ge}},\
  and\ \bibinfo {author} {\bibfnamefont {S.}~\bibnamefont {Hughes}},\ }\href
  {https://doi.org/10.1103/PhysRevA.92.053810} {\bibfield  {journal} {\bibinfo
  {journal} {Phys. Rev. A}\ }\textbf {\bibinfo {volume} {92}},\ \bibinfo
  {pages} {053810} (\bibinfo {year} {2015})}\BibitemShut {NoStop}%
\bibitem [{\citenamefont {Muljarov}\ and\ \citenamefont
  {Langbein}(2016)}]{Muljarov2016Phys.Rev.B}%
  \BibitemOpen
  \bibfield  {author} {\bibinfo {author} {\bibfnamefont {E.~A.}\ \bibnamefont
  {Muljarov}}\ and\ \bibinfo {author} {\bibfnamefont {W.}~\bibnamefont
  {Langbein}},\ }\href {https://doi.org/10.1103/PhysRevB.94.235438} {\bibfield
  {journal} {\bibinfo  {journal} {Phys. Rev. B}\ }\textbf {\bibinfo {volume}
  {94}},\ \bibinfo {pages} {235438} (\bibinfo {year} {2016})}\BibitemShut
  {NoStop}%
\bibitem [{\citenamefont {{El-Dardiry}}\ \emph {et~al.}(2012)\citenamefont
  {{El-Dardiry}}, \citenamefont {Faez},\ and\ \citenamefont
  {Lagendijk}}]{El-Dardiry2012Phys.Rev.B}%
  \BibitemOpen
  \bibfield  {author} {\bibinfo {author} {\bibfnamefont {R.~G.~S.}\
  \bibnamefont {{El-Dardiry}}}, \bibinfo {author} {\bibfnamefont
  {S.}~\bibnamefont {Faez}},\ and\ \bibinfo {author} {\bibfnamefont
  {A.}~\bibnamefont {Lagendijk}},\ }\href
  {https://doi.org/10.1103/PhysRevB.86.125132} {\bibfield  {journal} {\bibinfo
  {journal} {Phys. Rev. B}\ }\textbf {\bibinfo {volume} {86}},\ \bibinfo
  {pages} {125132} (\bibinfo {year} {2012})}\BibitemShut {NoStop}%
\bibitem [{\citenamefont {Vahala}(2003)}]{Vahala2003Nature}%
  \BibitemOpen
  \bibfield  {author} {\bibinfo {author} {\bibfnamefont {K.~J.}\ \bibnamefont
  {Vahala}},\ }\href {https://doi.org/10.1038/nature01939} {\bibfield
  {journal} {\bibinfo  {journal} {Nature}\ }\textbf {\bibinfo {volume} {424}},\
  \bibinfo {pages} {839} (\bibinfo {year} {2003})}\BibitemShut {NoStop}%
\bibitem [{\citenamefont {Aoki}(1983)}]{Aoki1983J.Phys.C:SolidStatePhys.}%
  \BibitemOpen
  \bibfield  {author} {\bibinfo {author} {\bibfnamefont {H.}~\bibnamefont
  {Aoki}},\ }\href {https://doi.org/10.1088/0022-3719/16/6/007} {\bibfield
  {journal} {\bibinfo  {journal} {J. Phys. C: Solid State Phys.}\ }\textbf
  {\bibinfo {volume} {16}},\ \bibinfo {pages} {L205} (\bibinfo {year}
  {1983})}\BibitemShut {NoStop}%
\bibitem [{\citenamefont {Janssen}(1998)}]{Janssen1998Phys.Rep.}%
  \BibitemOpen
  \bibfield  {author} {\bibinfo {author} {\bibfnamefont {M.}~\bibnamefont
  {Janssen}},\ }\href {https://doi.org/10.1016/S0370-1573(97)00050-1}
  {\bibfield  {journal} {\bibinfo  {journal} {Phys. Rep.}\ }\textbf {\bibinfo
  {volume} {295}},\ \bibinfo {pages} {1} (\bibinfo {year} {1998})}\BibitemShut
  {NoStop}%
\bibitem [{\citenamefont {Faez}\ \emph {et~al.}(2009)\citenamefont {Faez},
  \citenamefont {Strybulevych}, \citenamefont {Page}, \citenamefont
  {Lagendijk},\ and\ \citenamefont {{van Tiggelen}}}]{Faez2009Phys.Rev.Lett.}%
  \BibitemOpen
  \bibfield  {author} {\bibinfo {author} {\bibfnamefont {S.}~\bibnamefont
  {Faez}}, \bibinfo {author} {\bibfnamefont {A.}~\bibnamefont {Strybulevych}},
  \bibinfo {author} {\bibfnamefont {J.~H.}\ \bibnamefont {Page}}, \bibinfo
  {author} {\bibfnamefont {A.}~\bibnamefont {Lagendijk}},\ and\ \bibinfo
  {author} {\bibfnamefont {B.~A.}\ \bibnamefont {{van Tiggelen}}},\ }\href
  {https://doi.org/10.1103/PhysRevLett.103.155703} {\bibfield  {journal}
  {\bibinfo  {journal} {Phys. Rev. Lett.}\ }\textbf {\bibinfo {volume} {103}},\
  \bibinfo {pages} {155703} (\bibinfo {year} {2009})}\BibitemShut {NoStop}%
\bibitem [{\citenamefont {Abrahams}\ \emph {et~al.}(1979)\citenamefont
  {Abrahams}, \citenamefont {Anderson}, \citenamefont {Licciardello},\ and\
  \citenamefont {Ramakrishnan}}]{Abrahams1979Phys.Rev.Lett.}%
  \BibitemOpen
  \bibfield  {author} {\bibinfo {author} {\bibfnamefont {E.}~\bibnamefont
  {Abrahams}}, \bibinfo {author} {\bibfnamefont {P.~W.}\ \bibnamefont
  {Anderson}}, \bibinfo {author} {\bibfnamefont {D.~C.}\ \bibnamefont
  {Licciardello}},\ and\ \bibinfo {author} {\bibfnamefont {T.~V.}\ \bibnamefont
  {Ramakrishnan}},\ }\href {https://doi.org/10.1103/PhysRevLett.42.673}
  {\bibfield  {journal} {\bibinfo  {journal} {Phys. Rev. Lett.}\ }\textbf
  {\bibinfo {volume} {42}},\ \bibinfo {pages} {673} (\bibinfo {year}
  {1979})}\BibitemShut {NoStop}%
\bibitem [{\citenamefont {Sheng}(2006)}]{Sheng2006book}%
  \BibitemOpen
  \bibfield  {author} {\bibinfo {author} {\bibfnamefont {P.}~\bibnamefont
  {Sheng}},\ }\href@noop {} {\emph {\bibinfo {title} {Introduction to Wave
  Scattering, Localization, and Mesoscopic Phenomena}}},\ \bibinfo {edition}
  {2nd}\ ed.,\ \bibinfo {series} {Springer Series in Materials Science}\
  No.~\bibinfo {number} {88}\ (\bibinfo  {publisher} {{Springer}},\ \bibinfo
  {address} {{Berlin; New York}},\ \bibinfo {year} {2006})\BibitemShut
  {NoStop}%
\bibitem [{\citenamefont {Slevin}\ \emph {et~al.}(2001)\citenamefont {Slevin},
  \citenamefont {Marko{\v s}},\ and\ \citenamefont
  {Ohtsuki}}]{Slevin2001Phys.Rev.Lett.}%
  \BibitemOpen
  \bibfield  {author} {\bibinfo {author} {\bibfnamefont {K.}~\bibnamefont
  {Slevin}}, \bibinfo {author} {\bibfnamefont {P.}~\bibnamefont {Marko{\v
  s}}},\ and\ \bibinfo {author} {\bibfnamefont {T.}~\bibnamefont {Ohtsuki}},\
  }\href {https://doi.org/10.1103/PhysRevLett.86.3594} {\bibfield  {journal}
  {\bibinfo  {journal} {Phys. Rev. Lett.}\ }\textbf {\bibinfo {volume} {86}},\
  \bibinfo {pages} {3594} (\bibinfo {year} {2001})}\BibitemShut {NoStop}%
\bibitem [{\citenamefont {Bragg}\ and\ \citenamefont
  {Williams}(1934)}]{Bragg1934Proc.R.Soc.Lond.A}%
  \BibitemOpen
  \bibfield  {author} {\bibinfo {author} {\bibfnamefont {W.~L.}\ \bibnamefont
  {Bragg}}\ and\ \bibinfo {author} {\bibfnamefont {E.~J.}\ \bibnamefont
  {Williams}},\ }\href {https://doi.org/10.1098/rspa.1934.0132} {\bibfield
  {journal} {\bibinfo  {journal} {Proc. R. Soc. Lond. A}\ }\textbf {\bibinfo
  {volume} {145}},\ \bibinfo {pages} {699} (\bibinfo {year}
  {1934})}\BibitemShut {NoStop}%
\bibitem [{\citenamefont {Bethe}(1935)}]{Bethe1935Proc.R.Soc.Lond.A}%
  \BibitemOpen
  \bibfield  {author} {\bibinfo {author} {\bibfnamefont {H.~A.}\ \bibnamefont
  {Bethe}},\ }\href {https://doi.org/10.1098/rspa.1935.0122} {\bibfield
  {journal} {\bibinfo  {journal} {Proc. R. Soc. Lond. A}\ }\textbf {\bibinfo
  {volume} {150}},\ \bibinfo {pages} {552} (\bibinfo {year}
  {1935})}\BibitemShut {NoStop}%
\bibitem [{\citenamefont {Grahn}(1995)}]{Grahn1995book}%
  \BibitemOpen
  \bibinfo {editor} {\bibfnamefont {H.~T.}\ \bibnamefont {Grahn}},\ ed.,\
  \href@noop {} {\emph {\bibinfo {title} {Semiconductor Superlattices: Growth
  and Electronic Properties}}}\ (\bibinfo  {publisher} {{World Scientific}},\
  \bibinfo {address} {{Singapore ; River Edge, NJ}},\ \bibinfo {year}
  {1995})\BibitemShut {NoStop}%
\bibitem [{\citenamefont {Ando}\ \emph {et~al.}(1982)\citenamefont {Ando},
  \citenamefont {Fowler},\ and\ \citenamefont {Stern}}]{Ando1982Rev.Mod.Phys.}%
  \BibitemOpen
  \bibfield  {author} {\bibinfo {author} {\bibfnamefont {T.}~\bibnamefont
  {Ando}}, \bibinfo {author} {\bibfnamefont {A.~B.}\ \bibnamefont {Fowler}},\
  and\ \bibinfo {author} {\bibfnamefont {F.}~\bibnamefont {Stern}},\ }\href
  {https://doi.org/10.1103/RevModPhys.54.437} {\bibfield  {journal} {\bibinfo
  {journal} {Rev. Mod. Phys.}\ }\textbf {\bibinfo {volume} {54}},\ \bibinfo
  {pages} {437} (\bibinfo {year} {1982})}\BibitemShut {NoStop}%
\bibitem [{1()}]{1}%
  \BibitemOpen
  \href@noop {} {}\bibinfo {note} {Termed SAW-likeness
  coefficient.}\BibitemShut {Stop}%
\bibitem [{\citenamefont {Nardi}\ \emph {et~al.}(2009)\citenamefont {Nardi},
  \citenamefont {Banfi}, \citenamefont {Giannetti}, \citenamefont {Revaz},
  \citenamefont {Ferrini},\ and\ \citenamefont
  {Parmigiani}}]{Nardi2009Phys.Rev.B}%
  \BibitemOpen
  \bibfield  {author} {\bibinfo {author} {\bibfnamefont {D.}~\bibnamefont
  {Nardi}}, \bibinfo {author} {\bibfnamefont {F.}~\bibnamefont {Banfi}},
  \bibinfo {author} {\bibfnamefont {C.}~\bibnamefont {Giannetti}}, \bibinfo
  {author} {\bibfnamefont {B.}~\bibnamefont {Revaz}}, \bibinfo {author}
  {\bibfnamefont {G.}~\bibnamefont {Ferrini}},\ and\ \bibinfo {author}
  {\bibfnamefont {F.}~\bibnamefont {Parmigiani}},\ }\href
  {https://doi.org/10.1103/PhysRevB.80.104119} {\bibfield  {journal} {\bibinfo
  {journal} {Phys. Rev. B}\ }\textbf {\bibinfo {volume} {80}},\ \bibinfo
  {pages} {104119} (\bibinfo {year} {2009})}\BibitemShut {NoStop}%
\bibitem [{\citenamefont {Huang}\ and\ \citenamefont
  {Lee}(2012)}]{Huang2012Phys.Rev.B}%
  \BibitemOpen
  \bibfield  {author} {\bibinfo {author} {\bibfnamefont {B.}~\bibnamefont
  {Huang}}\ and\ \bibinfo {author} {\bibfnamefont {H.}~\bibnamefont {Lee}},\
  }\href {https://doi.org/10.1103/PhysRevB.86.245406} {\bibfield  {journal}
  {\bibinfo  {journal} {Phys. Rev. B}\ }\textbf {\bibinfo {volume} {86}},\
  \bibinfo {pages} {245406} (\bibinfo {year} {2012})}\BibitemShut {NoStop}%
\bibitem [{\citenamefont {Kratzer}\ and\ \citenamefont
  {Neugebauer}(2019)}]{Kratzer2019Front.Chem.}%
  \BibitemOpen
  \bibfield  {author} {\bibinfo {author} {\bibfnamefont {P.}~\bibnamefont
  {Kratzer}}\ and\ \bibinfo {author} {\bibfnamefont {J.}~\bibnamefont
  {Neugebauer}},\ }\href {https://doi.org/10.3389/fchem.2019.00106} {\bibfield
  {journal} {\bibinfo  {journal} {Front. Chem.}\ }\textbf {\bibinfo {volume}
  {7}},\ \bibinfo {pages} {106} (\bibinfo {year} {2019})}\BibitemShut {NoStop}%
\bibitem [{\citenamefont {Kresse}\ and\ \citenamefont
  {Furthm{\"u}ller}(1996)}]{Kresse1996ComputationalMaterialsScience}%
  \BibitemOpen
  \bibfield  {author} {\bibinfo {author} {\bibfnamefont {G.}~\bibnamefont
  {Kresse}}\ and\ \bibinfo {author} {\bibfnamefont {J.}~\bibnamefont
  {Furthm{\"u}ller}},\ }\href {https://doi.org/10.1016/0927-0256(96)00008-0}
  {\bibfield  {journal} {\bibinfo  {journal} {Computational Materials Science}\
  }\textbf {\bibinfo {volume} {6}},\ \bibinfo {pages} {15} (\bibinfo {year}
  {1996})}\BibitemShut {NoStop}%
\bibitem [{\citenamefont {Ho}\ \emph {et~al.}(1994)\citenamefont {Ho},
  \citenamefont {Chan}, \citenamefont {Soukoulis}, \citenamefont {Biswas},\
  and\ \citenamefont {Sigalas}}]{Ho1994SolidStateCommun.}%
  \BibitemOpen
  \bibfield  {author} {\bibinfo {author} {\bibfnamefont {K.~M.}\ \bibnamefont
  {Ho}}, \bibinfo {author} {\bibfnamefont {C.~T.}\ \bibnamefont {Chan}},
  \bibinfo {author} {\bibfnamefont {C.~M.}\ \bibnamefont {Soukoulis}}, \bibinfo
  {author} {\bibfnamefont {R.}~\bibnamefont {Biswas}},\ and\ \bibinfo {author}
  {\bibfnamefont {M.}~\bibnamefont {Sigalas}},\ }\href
  {https://doi.org/10.1016/0038-1098(94)90202-X} {\bibfield  {journal}
  {\bibinfo  {journal} {Solid State Commun.}\ }\textbf {\bibinfo {volume}
  {89}},\ \bibinfo {pages} {413} (\bibinfo {year} {1994})}\BibitemShut
  {NoStop}%
\bibitem [{\citenamefont {Devashish}\ \emph {et~al.}(2019)\citenamefont
  {Devashish}, \citenamefont {Ojambati}, \citenamefont {Hasan}, \citenamefont
  {{van der Vegt}},\ and\ \citenamefont {Vos}}]{Devashish2019Phys.Rev.B}%
  \BibitemOpen
  \bibfield  {author} {\bibinfo {author} {\bibfnamefont {D.}~\bibnamefont
  {Devashish}}, \bibinfo {author} {\bibfnamefont {O.~S.}\ \bibnamefont
  {Ojambati}}, \bibinfo {author} {\bibfnamefont {S.~B.}\ \bibnamefont {Hasan}},
  \bibinfo {author} {\bibfnamefont {J.~J.~W.}\ \bibnamefont {{van der Vegt}}},\
  and\ \bibinfo {author} {\bibfnamefont {W.~L.}\ \bibnamefont {Vos}},\ }\href
  {https://doi.org/10.1103/PhysRevB.99.075112} {\bibfield  {journal} {\bibinfo
  {journal} {Phys. Rev. B}\ }\textbf {\bibinfo {volume} {99}},\ \bibinfo
  {pages} {075112} (\bibinfo {year} {2019})}\BibitemShut {NoStop}%
\bibitem [{\citenamefont {Johnson}\ and\ \citenamefont
  {Joannopoulos}(2001)}]{Johnson2001Opt.Express}%
  \BibitemOpen
  \bibfield  {author} {\bibinfo {author} {\bibfnamefont {S.}~\bibnamefont
  {Johnson}}\ and\ \bibinfo {author} {\bibfnamefont {J.}~\bibnamefont
  {Joannopoulos}},\ }\href {https://doi.org/10.1364/OE.8.000173} {\bibfield
  {journal} {\bibinfo  {journal} {Opt. Express}\ }\textbf {\bibinfo {volume}
  {8}},\ \bibinfo {pages} {173} (\bibinfo {year} {2001})}\BibitemShut {NoStop}%
\end{thebibliography}%


\begin{thebibliography}{19}%
\makeatletter
\providecommand \@ifxundefined [1]{%
 \@ifx{#1\undefined}
}%
\providecommand \@ifnum [1]{%
 \ifnum #1\expandafter \@firstoftwo
 \else \expandafter \@secondoftwo
 \fi
}%
\providecommand \@ifx [1]{%
 \ifx #1\expandafter \@firstoftwo
 \else \expandafter \@secondoftwo
 \fi
}%
\providecommand \natexlab [1]{#1}%
\providecommand \enquote  [1]{``#1''}%
\providecommand \bibnamefont  [1]{#1}%
\providecommand \bibfnamefont [1]{#1}%
\providecommand \citenamefont [1]{#1}%
\providecommand \href@noop [0]{\@secondoftwo}%
\providecommand \href [0]{\begingroup \@sanitize@url \@href}%
\providecommand \@href[1]{\@@startlink{#1}\@@href}%
\providecommand \@@href[1]{\endgroup#1\@@endlink}%
\providecommand \@sanitize@url [0]{\catcode `\\12\catcode `\$12\catcode
  `\&12\catcode `\#12\catcode `\^12\catcode `\_12\catcode `\%12\relax}%
\providecommand \@@startlink[1]{}%
\providecommand \@@endlink[0]{}%
\providecommand \url  [0]{\begingroup\@sanitize@url \@url }%
\providecommand \@url [1]{\endgroup\@href {#1}{\urlprefix }}%
\providecommand \urlprefix  [0]{URL }%
\providecommand \Eprint [0]{\href }%
\providecommand \doibase [0]{https://doi.org/}%
\providecommand \selectlanguage [0]{\@gobble}%
\providecommand \bibinfo  [0]{\@secondoftwo}%
\providecommand \bibfield  [0]{\@secondoftwo}%
\providecommand \translation [1]{[#1]}%
\providecommand \BibitemOpen [0]{}%
\providecommand \bibitemStop [0]{}%
\providecommand \bibitemNoStop [0]{.\EOS\space}%
\providecommand \EOS [0]{\spacefactor3000\relax}%
\providecommand \BibitemShut  [1]{\csname bibitem#1\endcsname}%
\let\auto@bib@innerbib\@empty
\bibitem [{\citenamefont {Ku}\ \emph {et~al.}(2010)\citenamefont {Ku},
  \citenamefont {Berlijn},\ and\ \citenamefont {Lee}}]{Ku2010Phys.Rev.Lett.}%
  \BibitemOpen
  \bibfield  {author} {\bibinfo {author} {\bibfnamefont {W.}~\bibnamefont
  {Ku}}, \bibinfo {author} {\bibfnamefont {T.}~\bibnamefont {Berlijn}},\ and\
  \bibinfo {author} {\bibfnamefont {C.-C.}\ \bibnamefont {Lee}},\ }\href
  {https://doi.org/10.1103/PhysRevLett.104.216401} {\bibfield  {journal}
  {\bibinfo  {journal} {Phys. Rev. Lett.}\ }\textbf {\bibinfo {volume} {104}},\
  \bibinfo {pages} {216401} (\bibinfo {year} {2010})}\BibitemShut {NoStop}%
\bibitem [{\citenamefont {Mayo}\ \emph {et~al.}(2020)\citenamefont {Mayo},
  \citenamefont {Yndurain},\ and\ \citenamefont
  {Soler}}]{Mayo2020J.Phys.Condens.Matter}%
  \BibitemOpen
  \bibfield  {author} {\bibinfo {author} {\bibfnamefont {S.~G.}\ \bibnamefont
  {Mayo}}, \bibinfo {author} {\bibfnamefont {F.}~\bibnamefont {Yndurain}},\
  and\ \bibinfo {author} {\bibfnamefont {J.~M.}\ \bibnamefont {Soler}},\ }\href
  {https://doi.org/10.1088/1361-648X/ab6e8e} {\bibfield  {journal} {\bibinfo
  {journal} {J. Phys. Condens. Matter}\ }\textbf {\bibinfo {volume} {32}},\
  \bibinfo {pages} {205902} (\bibinfo {year} {2020})}\BibitemShut {NoStop}%
\bibitem [{\citenamefont {Aoki}(1983)}]{Aoki1983J.Phys.C:SolidStatePhys.}%
  \BibitemOpen
  \bibfield  {author} {\bibinfo {author} {\bibfnamefont {H.}~\bibnamefont
  {Aoki}},\ }\href {https://doi.org/10.1088/0022-3719/16/6/007} {\bibfield
  {journal} {\bibinfo  {journal} {J. Phys. C: Solid State Phys.}\ }\textbf
  {\bibinfo {volume} {16}},\ \bibinfo {pages} {L205} (\bibinfo {year}
  {1983})}\BibitemShut {NoStop}%
\bibitem [{\citenamefont {Janssen}(1998)}]{Janssen1998Phys.Rep.}%
  \BibitemOpen
  \bibfield  {author} {\bibinfo {author} {\bibfnamefont {M.}~\bibnamefont
  {Janssen}},\ }\href {https://doi.org/10.1016/S0370-1573(97)00050-1}
  {\bibfield  {journal} {\bibinfo  {journal} {Phys. Rep.}\ }\textbf {\bibinfo
  {volume} {295}},\ \bibinfo {pages} {1} (\bibinfo {year} {1998})}\BibitemShut
  {NoStop}%
\bibitem [{\citenamefont {Faez}\ \emph {et~al.}(2009)\citenamefont {Faez},
  \citenamefont {Strybulevych}, \citenamefont {Page}, \citenamefont
  {Lagendijk},\ and\ \citenamefont {{van Tiggelen}}}]{Faez2009Phys.Rev.Lett.}%
  \BibitemOpen
  \bibfield  {author} {\bibinfo {author} {\bibfnamefont {S.}~\bibnamefont
  {Faez}}, \bibinfo {author} {\bibfnamefont {A.}~\bibnamefont {Strybulevych}},
  \bibinfo {author} {\bibfnamefont {J.~H.}\ \bibnamefont {Page}}, \bibinfo
  {author} {\bibfnamefont {A.}~\bibnamefont {Lagendijk}},\ and\ \bibinfo
  {author} {\bibfnamefont {B.~A.}\ \bibnamefont {{van Tiggelen}}},\ }\href
  {https://doi.org/10.1103/PhysRevLett.103.155703} {\bibfield  {journal}
  {\bibinfo  {journal} {Phys. Rev. Lett.}\ }\textbf {\bibinfo {volume} {103}},\
  \bibinfo {pages} {155703} (\bibinfo {year} {2009})}\BibitemShut {NoStop}%
\bibitem [{\citenamefont {Hurley}\ and\ \citenamefont
  {Rickard}(2009)}]{Hurley2009IEEETrans.Inform.Theory}%
  \BibitemOpen
  \bibfield  {author} {\bibinfo {author} {\bibfnamefont {N.}~\bibnamefont
  {Hurley}}\ and\ \bibinfo {author} {\bibfnamefont {S.}~\bibnamefont
  {Rickard}},\ }\href {https://doi.org/10.1109/TIT.2009.2027527} {\bibfield
  {journal} {\bibinfo  {journal} {IEEE Trans. Inform. Theory}\ }\textbf
  {\bibinfo {volume} {55}},\ \bibinfo {pages} {4723} (\bibinfo {year}
  {2009})}\BibitemShut {NoStop}%
\bibitem [{\citenamefont {{El-Dardiry}}\ \emph {et~al.}(2012)\citenamefont
  {{El-Dardiry}}, \citenamefont {Faez},\ and\ \citenamefont
  {Lagendijk}}]{El-Dardiry2012Phys.Rev.B}%
  \BibitemOpen
  \bibfield  {author} {\bibinfo {author} {\bibfnamefont {R.~G.~S.}\
  \bibnamefont {{El-Dardiry}}}, \bibinfo {author} {\bibfnamefont
  {S.}~\bibnamefont {Faez}},\ and\ \bibinfo {author} {\bibfnamefont
  {A.}~\bibnamefont {Lagendijk}},\ }\href
  {https://doi.org/10.1103/PhysRevB.86.125132} {\bibfield  {journal} {\bibinfo
  {journal} {Phys. Rev. B}\ }\textbf {\bibinfo {volume} {86}},\ \bibinfo
  {pages} {125132} (\bibinfo {year} {2012})}\BibitemShut {NoStop}%
\bibitem [{\citenamefont {Arceri}\ and\ \citenamefont
  {Corwin}(2020)}]{Arceri2020Phys.Rev.Lett.}%
  \BibitemOpen
  \bibfield  {author} {\bibinfo {author} {\bibfnamefont {F.}~\bibnamefont
  {Arceri}}\ and\ \bibinfo {author} {\bibfnamefont {E.~I.}\ \bibnamefont
  {Corwin}},\ }\href {https://doi.org/10.1103/PhysRevLett.124.238002}
  {\bibfield  {journal} {\bibinfo  {journal} {Phys. Rev. Lett.}\ }\textbf
  {\bibinfo {volume} {124}},\ \bibinfo {pages} {238002} (\bibinfo {year}
  {2020})}\BibitemShut {NoStop}%
\bibitem [{\citenamefont {Pashartis}\ and\ \citenamefont
  {Rubel}(2017)}]{Pashartis2017Phys.Rev.B}%
  \BibitemOpen
  \bibfield  {author} {\bibinfo {author} {\bibfnamefont {C.}~\bibnamefont
  {Pashartis}}\ and\ \bibinfo {author} {\bibfnamefont {O.}~\bibnamefont
  {Rubel}},\ }\href {https://doi.org/10.1103/PhysRevB.96.155209} {\bibfield
  {journal} {\bibinfo  {journal} {Phys. Rev. B}\ }\textbf {\bibinfo {volume}
  {96}},\ \bibinfo {pages} {155209} (\bibinfo {year} {2017})}\BibitemShut
  {NoStop}%
\bibitem [{\citenamefont {Kratzer}\ and\ \citenamefont
  {Neugebauer}(2019)}]{Kratzer2019Front.Chem.}%
  \BibitemOpen
  \bibfield  {author} {\bibinfo {author} {\bibfnamefont {P.}~\bibnamefont
  {Kratzer}}\ and\ \bibinfo {author} {\bibfnamefont {J.}~\bibnamefont
  {Neugebauer}},\ }\href {https://doi.org/10.3389/fchem.2019.00106} {\bibfield
  {journal} {\bibinfo  {journal} {Front. Chem.}\ }\textbf {\bibinfo {volume}
  {7}},\ \bibinfo {pages} {106} (\bibinfo {year} {2019})}\BibitemShut {NoStop}%
\bibitem [{\citenamefont {Kresse}\ and\ \citenamefont
  {Furthm{\"u}ller}(1996)}]{Kresse1996ComputationalMaterialsScience}%
  \BibitemOpen
  \bibfield  {author} {\bibinfo {author} {\bibfnamefont {G.}~\bibnamefont
  {Kresse}}\ and\ \bibinfo {author} {\bibfnamefont {J.}~\bibnamefont
  {Furthm{\"u}ller}},\ }\href {https://doi.org/10.1016/0927-0256(96)00008-0}
  {\bibfield  {journal} {\bibinfo  {journal} {Computational Materials Science}\
  }\textbf {\bibinfo {volume} {6}},\ \bibinfo {pages} {15} (\bibinfo {year}
  {1996})}\BibitemShut {NoStop}%
\bibitem [{\citenamefont {Perdew}\ \emph {et~al.}(1996)\citenamefont {Perdew},
  \citenamefont {Burke},\ and\ \citenamefont
  {Ernzerhof}}]{Perdew1996Phys.Rev.Lett.}%
  \BibitemOpen
  \bibfield  {author} {\bibinfo {author} {\bibfnamefont {J.~P.}\ \bibnamefont
  {Perdew}}, \bibinfo {author} {\bibfnamefont {K.}~\bibnamefont {Burke}},\ and\
  \bibinfo {author} {\bibfnamefont {M.}~\bibnamefont {Ernzerhof}},\ }\href
  {https://doi.org/10.1103/PhysRevLett.77.3865} {\bibfield  {journal} {\bibinfo
   {journal} {Phys. Rev. Lett.}\ }\textbf {\bibinfo {volume} {77}},\ \bibinfo
  {pages} {3865} (\bibinfo {year} {1996})}\BibitemShut {NoStop}%
\bibitem [{\citenamefont {Kresse}\ and\ \citenamefont
  {Joubert}(1999)}]{Kresse1999Phys.Rev.B}%
  \BibitemOpen
  \bibfield  {author} {\bibinfo {author} {\bibfnamefont {G.}~\bibnamefont
  {Kresse}}\ and\ \bibinfo {author} {\bibfnamefont {D.}~\bibnamefont
  {Joubert}},\ }\href {https://doi.org/10.1103/PhysRevB.59.1758} {\bibfield
  {journal} {\bibinfo  {journal} {Phys. Rev. B}\ }\textbf {\bibinfo {volume}
  {59}},\ \bibinfo {pages} {1758} (\bibinfo {year} {1999})}\BibitemShut
  {NoStop}%
\bibitem [{\citenamefont {Hillebrand}\ \emph {et~al.}(2003)\citenamefont
  {Hillebrand}, \citenamefont {Senz}, \citenamefont {Hergert},\ and\
  \citenamefont {G{\"o}sele}}]{Hillebrand2003J.Appl.Phys.}%
  \BibitemOpen
  \bibfield  {author} {\bibinfo {author} {\bibfnamefont {R.}~\bibnamefont
  {Hillebrand}}, \bibinfo {author} {\bibfnamefont {S.}~\bibnamefont {Senz}},
  \bibinfo {author} {\bibfnamefont {W.}~\bibnamefont {Hergert}},\ and\ \bibinfo
  {author} {\bibfnamefont {U.}~\bibnamefont {G{\"o}sele}},\ }\href
  {https://doi.org/10.1063/1.1593796} {\bibfield  {journal} {\bibinfo
  {journal} {J. Appl. Phys.}\ }\textbf {\bibinfo {volume} {94}},\ \bibinfo
  {pages} {2758} (\bibinfo {year} {2003})}\BibitemShut {NoStop}%
\bibitem [{\citenamefont {Ho}\ \emph {et~al.}(1994)\citenamefont {Ho},
  \citenamefont {Chan}, \citenamefont {Soukoulis}, \citenamefont {Biswas},\
  and\ \citenamefont {Sigalas}}]{Ho1994SolidStateCommun.}%
  \BibitemOpen
  \bibfield  {author} {\bibinfo {author} {\bibfnamefont {K.~M.}\ \bibnamefont
  {Ho}}, \bibinfo {author} {\bibfnamefont {C.~T.}\ \bibnamefont {Chan}},
  \bibinfo {author} {\bibfnamefont {C.~M.}\ \bibnamefont {Soukoulis}}, \bibinfo
  {author} {\bibfnamefont {R.}~\bibnamefont {Biswas}},\ and\ \bibinfo {author}
  {\bibfnamefont {M.}~\bibnamefont {Sigalas}},\ }\href
  {https://doi.org/10.1016/0038-1098(94)90202-X} {\bibfield  {journal}
  {\bibinfo  {journal} {Solid State Commun.}\ }\textbf {\bibinfo {volume}
  {89}},\ \bibinfo {pages} {413} (\bibinfo {year} {1994})}\BibitemShut
  {NoStop}%
\bibitem [{\citenamefont {Leistikow}\ \emph {et~al.}(2011)\citenamefont
  {Leistikow}, \citenamefont {Mosk}, \citenamefont {Yeganegi}, \citenamefont
  {Huisman}, \citenamefont {Lagendijk},\ and\ \citenamefont
  {Vos}}]{Leistikow2011Phys.Rev.Lett.}%
  \BibitemOpen
  \bibfield  {author} {\bibinfo {author} {\bibfnamefont {M.~D.}\ \bibnamefont
  {Leistikow}}, \bibinfo {author} {\bibfnamefont {A.~P.}\ \bibnamefont {Mosk}},
  \bibinfo {author} {\bibfnamefont {E.}~\bibnamefont {Yeganegi}}, \bibinfo
  {author} {\bibfnamefont {S.~R.}\ \bibnamefont {Huisman}}, \bibinfo {author}
  {\bibfnamefont {A.}~\bibnamefont {Lagendijk}},\ and\ \bibinfo {author}
  {\bibfnamefont {W.~L.}\ \bibnamefont {Vos}},\ }\href
  {https://doi.org/10.1103/PhysRevLett.107.193903} {\bibfield  {journal}
  {\bibinfo  {journal} {Phys. Rev. Lett.}\ }\textbf {\bibinfo {volume} {107}},\
  \bibinfo {pages} {193903} (\bibinfo {year} {2011})}\BibitemShut {NoStop}%
\bibitem [{\citenamefont {Johnson}\ and\ \citenamefont
  {Joannopoulos}(2001)}]{Johnson2001Opt.Express}%
  \BibitemOpen
  \bibfield  {author} {\bibinfo {author} {\bibfnamefont {S.}~\bibnamefont
  {Johnson}}\ and\ \bibinfo {author} {\bibfnamefont {J.}~\bibnamefont
  {Joannopoulos}},\ }\href {https://doi.org/10.1364/OE.8.000173} {\bibfield
  {journal} {\bibinfo  {journal} {Opt. Express}\ }\textbf {\bibinfo {volume}
  {8}},\ \bibinfo {pages} {173} (\bibinfo {year} {2001})}\BibitemShut {NoStop}%
\bibitem [{\citenamefont {Setyawan}\ and\ \citenamefont
  {Curtarolo}(2010)}]{Setyawan2010Comput.Mater.Sci.}%
  \BibitemOpen
  \bibfield  {author} {\bibinfo {author} {\bibfnamefont {W.}~\bibnamefont
  {Setyawan}}\ and\ \bibinfo {author} {\bibfnamefont {S.}~\bibnamefont
  {Curtarolo}},\ }\href {https://doi.org/10.1016/j.commatsci.2010.05.010}
  {\bibfield  {journal} {\bibinfo  {journal} {Comput. Mater. Sci.}\ }\textbf
  {\bibinfo {volume} {49}},\ \bibinfo {pages} {299} (\bibinfo {year}
  {2010})}\BibitemShut {NoStop}%
\bibitem [{\citenamefont {Woldering}\ \emph {et~al.}(2014)\citenamefont
  {Woldering}, \citenamefont {Mosk},\ and\ \citenamefont
  {Vos}}]{Woldering2014Phys.Rev.B}%
  \BibitemOpen
  \bibfield  {author} {\bibinfo {author} {\bibfnamefont {L.~A.}\ \bibnamefont
  {Woldering}}, \bibinfo {author} {\bibfnamefont {A.~P.}\ \bibnamefont
  {Mosk}},\ and\ \bibinfo {author} {\bibfnamefont {W.~L.}\ \bibnamefont
  {Vos}},\ }\href {https://doi.org/10.1103/PhysRevB.90.115140} {\bibfield
  {journal} {\bibinfo  {journal} {Phys. Rev. B}\ }\textbf {\bibinfo {volume}
  {90}},\ \bibinfo {pages} {115140} (\bibinfo {year} {2014})}\BibitemShut
  {NoStop}%
\end{thebibliography}%

\makeatletter\@input{xx.tex}\makeatother 
\end{document}


\title{Supplemental material}
\maketitle

\section{Basic considerations}
The larger the supercell gets, the more bands it contains in the given frequency range.
This effect is known as band-folding and it is a fundamental property of the Brillouin zone.
Several ways of "unfolding" the band structure have been proposed in the literature, see, \eg , Refs.~\cite{Ku2010Phys.Rev.Lett., Mayo2020J.Phys.Condens.Matter}.
Nevertheless, it is still very hard to draw direct relations between bands from different supercells as the band-folding effects seem to also depend on the confinement dimensionality of the bands.
Our approach circumvents this problem by not requiring such direct relations between the folded and unfolded bands.
Namely, in the $\VV^\alpha/\EE$ plots for large enough $N>N_0$, each band of the investigated $N$ supercell is clearly either above or below \textit{all} the bands with similar frequencies of the reference $N_0$ supercell.

Our method can be viewed as an extension of the multifractality techniques, which are usually formulated for scaling of the participation ratio~\cite{Aoki1983J.Phys.C:SolidStatePhys.,Janssen1998Phys.Rep., Faez2009Phys.Rev.Lett.}.
In the first step of our framework we also perform a similar scaling for the mode volume in Eq.~\eq{main-eq:vmodeecavD}.
However, by limiting oneself to only this scaling behavior, it would require very large supercell size $N$ to draw conclusive results for the whole spectrum, as described below and in Fig.~\ref{fig:mffailure}.
To complicate the matter even more, due to the band folding effects, translating the results obtained for this large $N$ to experimentally interesting supercells of moderate size would be extremely complicated, if at all possible.
By subsequently introducing the confinement energy as another analyzed quantity with different scaling exponent than the mode volume and the auxiliary power parameter $\alpha$, our method allows for determination of confinement throughout the whole spectrum for relatively low supercell sizes and, as described in the previous paragraph, free of complications related to band folding.

In our definition of $\Ecav$ in \eq{main-ecavdef} we choose the volume $\Vcav$ to be centered around the cavity, where the confinement is expected.
Nevertheless, this is not a strict requirement.
Choosing the volume $\Vcav$ at a different location within the supercell will still yield correct scaling results, albeit possibly affecting the convergence of the method and thus requiring larger supercell sizes $N, N_0$ to properly classify the whole spectrum.
It is however necessary, after $\Vcav$ is selected, to keep this volume constant throughout the scaling, so that the exponent $c-D$ appears in the scaling relation for $\EE$ in \eq{main-eq:vmodeecavD}, since this is crucial for being able to tune the values of the critical exponent $\kappa$ as needed.

The dimension of the supercell used to model the system may differ from the system dimensionality $D$ in some cases. 
For example, in 2D electronic materials $D=2$, but the charge density usually extends slightly into the third dimension and, moreover, can be zero in the 2D plane where the atomic nuclei are positioned.
This case is therefore traditionally modeled by a 3D supercell and thus the integration in Eqs.~\eq{main-mvdef} and~\eq{main-ecavdef} must be performed over its whole three-dimensional volume, \ie , the volumes $\Vsup$ and $\Vcav$ will be three-dimensional.
Nevertheless, since the structure of the material is still two-dimensional, the subsequent scaling analysis of confinement can be performed in $D=2$ dimensions with the third dimension kept constant, \ie, adding unit cells only along the planar material, as we did in the example of a hexagonal BN with N vacancy.

In this Letter we are only concerned with integer defect dimensionalities $d$ in systems of dimension $D\le 3$, as these are the most relevant for experiments. 
It is however straightforward to extend our technique to also include fractal defects with fractional dimensionalities and higher dimensional spaces.
One can directly apply the derivations in this Supplemental material to find the optimal values of the auxiliary power $\alpha$ in case of fractional $d$ or arbitrarily high $D$.
These possibilities further emphasize the generality of our scaling theory.

\section{Sparsity measures: Analogies of mode volume and participation ratio}
In our confinement analysis, we employ the mode volume $\Vmode$ as one of two key quantities, defined in Eq.~\eq{main-mvdef}.
This is, however, not the only possible choice, and a whole class of quantities can be substituted for $\Vmode$ in our confinement identification method.

Ref.~\cite{Hurley2009IEEETrans.Inform.Theory} is the first to rigorously define the so-called \textit{measures of sparsity}.
In the most simple terms, measures of sparsity are functions quantifying the inequality of distribution of values within a given set.
One such measure, for the set of values $\{m_1,\ldots,m_M\}\in\RR^M$, is a $pq$-mean:
\begin{equation}
\mu_{p,q} \colonequals \frac{\left(\frac{1}{M}\sum\limits_{m=1}^{M}c_j^p\right)^{\nicefrac{1}{p}}}{\left(\frac{1}{M}\sum\limits_{m=1}^{M}c_j^q\right)^{\nicefrac{1}{q}}}.
\label{pqmeandef}
\end{equation}
Ref.~\cite{Hurley2009IEEETrans.Inform.Theory} has shown that, for $p<q<\infty$, $\mu_{p,q}$ is indeed a viable measure of sparsity.

There is clear correspondence between the inequality of $W(\vb x)$ over the volume $\Vsup$ for a given band and its confinement: For a confined band, $W(\vb x)$ is very high in certain parts of $\Vsup$ but low everywhere else, while for an extended band $W(\vb x)$ is approximately equally distributed over the whole $\Vsup$.
To see this quantitatively, we write the straightforward continuous generalization of \eq{pqmeandef} to quantify the inequality of $W(\vb x)$ over the volume $\Vsup$ as
\begin{equation}
U_{p,q} \colonequals \frac{\left(\frac{1}{\Vsup}\int\limits_{\Vsup}W^p(\vb x)\dd{V}\right)^{\nicefrac{1}{p}}}{\left(\frac{1}{\Vsup}\int\limits_{\Vsup}W^q(\vb x)\dd{V}\right)^{\nicefrac{1}{q}}}.
\label{pqmeancont}
\end{equation}
Our normalized mode volume $\VV$, as defined by Eq.~\eq{main-venorm}, turns out to be a continuous version of the $pq$-mean with $p=1$ and $q=\infty$, \ie , $\VV=U_{1,\infty}$.
The fact that $q=\infty$ means that $\VV$ has slightly weaker properties than the sparsity measure defined by Ref.~\cite{Hurley2009IEEETrans.Inform.Theory}. 
This detail is, however, unimportant for our confinement identification purposes.

For the purpose of confinement analysis, one can substitute any other $pq$-mean $U_{p,q}$ for $\VV$ and our method will still work properly, albeit possibly with different scaling power than we obtained for $\VV$ in Eq.~\eq{main-eq:vmodeecavD}.
Specifically, substituting $p=1$, $q=2$ yields
\begin{equation}
U_{1,2} \colonequals \frac{\int\limits_{\Vsup}W(\vb x)\dd{V}}{\sqrt{\Vsup}\sqrt{\int\limits_{\Vsup}W^2(\vb x)\dd{V}}}.
\label{pdef}
\end{equation}
The quantity $U_{1,2}$ corresponds to the square root of normalized participation ratio:
\begin{equation}
U_{1,2}=\sqrt{\frac{P}{\Vsup}},
\label{ppr}
\end{equation}
where the participation ratio $P$~\cite{El-Dardiry2012Phys.Rev.B, Arceri2020Phys.Rev.Lett., Pashartis2017Phys.Rev.B} is given by
\begin{equation}
P \colonequals \frac{\left(\int_{\Vsup} W(\vb x)\dd{V}\right)^2}{\int_{\Vsup} W^2(\vb x)\dd{V}}.
\label{prdef}
\end{equation}
Finally, since the $pq$-mean is a non-negative dimensionless quantity, its various powers are equivalent in characterizing confinement and thus one can use $U_{1,2}^2$ instead of $U_{1,2}$.
This illustrates that, in our approach, instead of normalized mode volume $\VV$, normalized participation ratio can be used as well, similarly to any other $pq$-mean $U_{p,q}$.

\section{Failure of multifractality}
\begin{figure}
	\includegraphics[width=\linewidth]{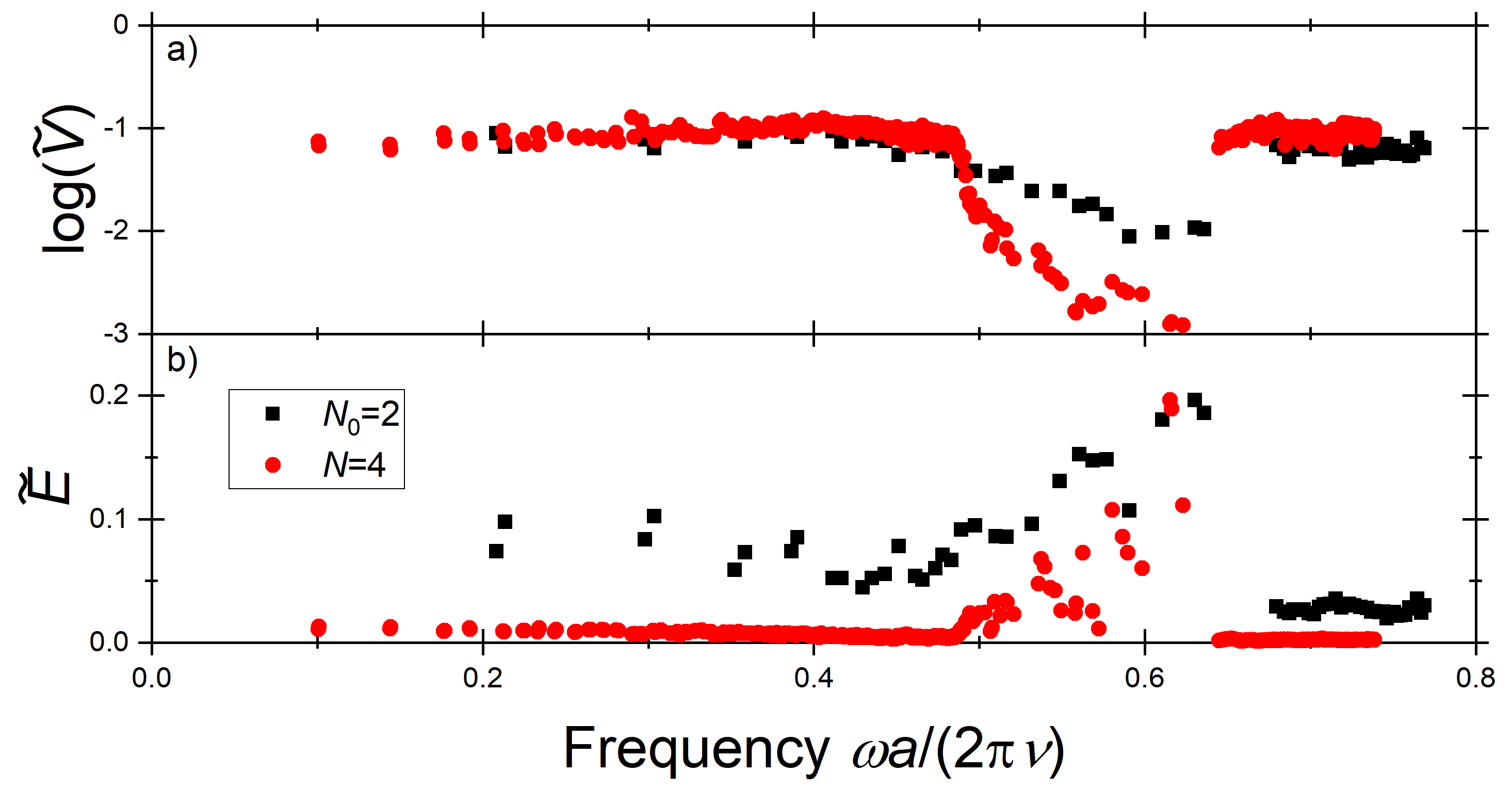}
	\caption{ Failure of the standard multifractality approach to identify confined bands for small supercell sizes. 
	Each band is represented by a point.
	(a) Scaling of the normalized mode volume. 
	(b) Scaling of the normalized confinement energy.
	For some bands the plotted values change significantly when scaling from $N_0=2$ to $N=4$ while for others they remain approximately constant.
	This variety of behaviors indicates that different confinement dimensionalities $c$ are present in the band spectrum, as predicted by the multifractality.
	Nevertheless, from each of these plots on its own, it is not possible to clearly assign confinement dimensionalities to specific bands and there is thus a need to move beyond the traditional multifractality method.
	One can, albeit, infer interesting information on confinement properties upon analyzing both plots simultaneously, which already brings us close to the essence of our technique of scaling the ratio of $\VV^\alpha/\EE$.
	}
	\label{fig:mffailure}
\end{figure}

In the main text we claim that our method surpasses multifractality analysis because it is able to identify bands in much smaller supercells than is possible by multifractality.
We illustrate this by applying the simple multifractality scaling to our example of 3D inverse woodpile photonic crystal with two linear defects used in the main text.
We analyze the $N=4$ supercell and use the $N_0=2$ supercell as a reference.
We scale either the mode volume $\VV$, in Fig.~\ref{fig:mffailure}(a), or the confinement energy $\EE$, in Fig.~\ref{fig:mffailure}(b).

From Fig.~\ref{fig:mffailure}(a), we observe that for some bands the value of $\VV$ decreases significantly as the supercell grows and for some it stays roughly constant. 
One would immediately expect the bands with constant $\VV$ to be extended, according to Eq.~\eq{main-eq:vmodeecavD}.
However, differentiating between the $c=2$ and $c=3$ bands is basically impossible.
Moreover, even the distinction between the "decreasing" and "constant" mode volume can be rather unclear, for example around the frequencies $\tilde\omega\approx0.5$.

Similarly, in Fig.~\ref{fig:mffailure}(b), for some bands the value of $\EE$ decreases significantly as the supercell grows and for some it stays roughly constant.
In this case, the distinction between the various confinement dimensionalities is even blurrier than for the mode volume.

As it is suggestively implied by the combined Fig.~\ref{fig:mffailure}, looking at the scaling of both $\VV$ and $\EE$ at the same time and comparing these yields more information than analyzing only one of them as does the traditional multifractality approach.
This already brings us very close to our confinement analysis method as described in the main text, which represents a systematic approach to the simultaneous scaling of both $\VV$ and $\EE$.

\section{Derivation of the scaling equation}
In a homogeneous $D$-dimensional system, waves propagate freely in all directions.
Adding a periodic structure with no defects may restrict some ways of propagation, but the waves will still be extended, \ie , they can propagate from any given unit cell to any other one.
Upon introducing a defect of dimensionality $d\le D$, some waves may couple to this defect and only propagate within it, thus never being able to achieve the non-defect unit cells.
A wave coupled to a defect of dimension $d$ has confinement dimensionality $c=D-d$, \ie , it is confined in $c$ dimensions.
This effect is very strongly observable upon scaling of the system size.
To analyze the confinement dimensionality of waves, one can utilize the scaling of mode volume and confinement energy.

Mathematically, the behavior of mode volume, defined by \eq{main-mvdef}, for such a wave is given by
\begin{equation}
\begin{aligned}
\Vmode &= AN^{d}+\OO(N^{d-1}) \\
&= AN^{D-c}+\OO(N^{D-c-1}),
\end{aligned}
\label{mvscaling}
\end{equation}
where $A$ is a constant independent of $N$.
In fact, $A$ is actually a functional depending on the permittivity distribution $\epsilon(\vb x)$, the wave-vector $\vb k$ and the frequency $\omega$.
Eq.~\eq{mvscaling} holds beyond the localization length of a given wave and expresses the fact that the mode volume of a confined wave can only grow within the geometrical constraints of the defect, with a small contribution of decaying waves in other directions.
Upon normalization by $\Vsup=N^D$, we obtain
\begin{equation}
\VV  = AN^{-c}+\OO(N^{-c-1}).
\label{vvscaling}
\end{equation}

Similarly, the normalized confinement energy, defined by \eq{main-ecavdef} and \eq{main-eq:vmodeecavD}, behaves as
\begin{equation}
\begin{aligned}
\EE &=  B N^{-d}+\OO(N^{-d-1}) \\
&=  BN^{c-D}+\OO(N^{c-D-1}),
\end{aligned}
\label{eescaling}
\end{equation}
where $B$ is a constant independent of $N$.
Analogously to $A$, $B$ is in fact a functional depending on the permittivity distribution $\epsilon(\vb x)$, the wave-vector $\vb k$ and the frequency $\omega$.
Eq.~\eq{eescaling} holds beyond the localization length of the given wave and expresses the fact that the confinement energy of a confined wave can only escape from the volume $V_0$ in ways that obey the geometrical constraints of the defect. 
There can also be additional, fast decaying, leakage expressed by the term $\OO(N^{c-D-1})$. 

We combine the Eqs.~\eq{vvscaling} and~\eq{eescaling} and, upon neglecting the big-$\OO$ contributions, we obtain for a wave with the confinement dimensionality $c$ in the limit $N\ra\infty$:
\begin{equation}
\frac{\VV^\alpha}{\EE} =\frac{\left(AN^{-c}\right)^\alpha}{BN^{c-D}} = CN^{-(\alpha+1) c+D},
\label{vveederivedscalinglim}
\end{equation}
where $C=A^\alpha/B$.
Eq.~\eq{vveederivedscalinglim} represents exactly our scaling relation in Eq.~\eq{main-eq:scaling}.

\section{Determining the auxiliary powers $\alpha$}
In our method, we are able to distinguish between the bands with $c< j$ and $c\ge j$ by tracking the behavior of $\VV^\alpha/\EE$ according to Eq.~\eq{main-eq:scaling}, as the supercell size changes from a smaller reference size $N_0$ to the investigated size $N$.
By varying $j$ over all the values $0< j \le D$, this approach allows us to completely classify all wave bands in the spectrum based on their confinement dimensionality.
For this approach to work, the power $\alpha$ has to be chosen so that $\kappa<0$ for all $c\ge j$ and $\kappa>0$ for all $c< j$, where $\kappa$ is given by Eq.~\eq{main-eq:kappa}.
This condition, however, still allows for some freedom in the choice of $\alpha$, which can be used to reduce the influence of the sub-leading orders to the Eq.~\eq{main-eq:scaling}.

These sub-leading orders are represented by the big-$\OO$ contributions in Eqs.~\eq{vvscaling} and~\eq{eescaling} and, especially for small supercells, can change the behavior of certain bands from what would be expected based on Eq.~\eq{main-eq:scaling}.
Since we determine the confinement dimensionality by analyzing the change of $\VV^\alpha/\EE$ with respect to the reference supercell, the bands most susceptible to the sub-leading order effects will be the bands where the smallest leading-order change is expected.
According to Eq.~\eq{main-eq:scaling}, these are the bands, for which $\kappa$ is close to zero, \ie , for every $j$, the bands with $c=j$ and $c=j-1$.
Because we generally do not know the contribution of the sub-leading orders to the overall scaling result, the most reasonable choice is to maximize the leading-order movement, \ie , that of the $c=j$ bands downwards and that of the $c=j-1$ bands upwards, at the same time.

Since $\kappa\colonequals-(\alpha+1) c+D$ is a linear function of $c$, the desired leading-order movements will be maximal if $\kappa=0$ at the middle point, \ie , for $c=j-1/2$. 
Thus, for each $0 < j \le D$, we solve the equation
\begin{equation}
-(\alpha+1)(j-\frac{1}{2})+D =0.
\label{eqalpha}
\end{equation}
The solution to \eq{eqalpha} is
\begin{equation}
\alpha = \frac{2D}{2j-1} - 1.
\label{alphavalues}
\end{equation}
By choosing $\alpha$ according to \eq{alphavalues} for each $D,j$, we ensure the effective minimization of the sub-leading order contributions to our scaling analysis and thus the best precision of our method.
The powers $\alpha$ obtained this way for each $j$ for $D=1,2,3$ are tabulated in Table~\ref{main-table:kappa}.

\section{Electronic band computation}
For our analysis of the 2D hexagonal BN layer with N vacancy, we calculated the band structure and charge densities using the density functional theory~\cite{Kratzer2019Front.Chem.} implemented in the VASP code v6.1.1~\cite{Kresse1996ComputationalMaterialsScience}.
We use the PBE exchange-correlation functional~\cite{Perdew1996Phys.Rev.Lett.} and represent the electron-ion interaction via the PAW potentials~\cite{Kresse1999Phys.Rev.B}, specifically via the B and N potential files recommended by VASP.
The cutoff energy was set to 550 eV.

We modeled the material using a 3D supercell with periodic boundary conditions, starting each geometry optimization with the vacuum layer of 12.56 $\mr{\mathring{A}}$ perpendicular to the material layer to eliminate the effects of periodic boundary conditions in this direction.
For geometry optimization and computation of the charge densities, we used $11\times11$ $\Gamma$-centered $k$-point grid.
Geometry optimization was performed via the conjugate gradient algorithm until the residual atomic forces were smaller than 10 $\mr{meV/\mathring{A}}$.
For the computation of the band structure in Fig.~\ref{main-fig:DFTconfinement}(a), we interpolated each segment of the high-symmetry path with 13 $k$-points, in addition to the two high-symmetry points.

\begin{figure}
	\includegraphics[width=\linewidth]{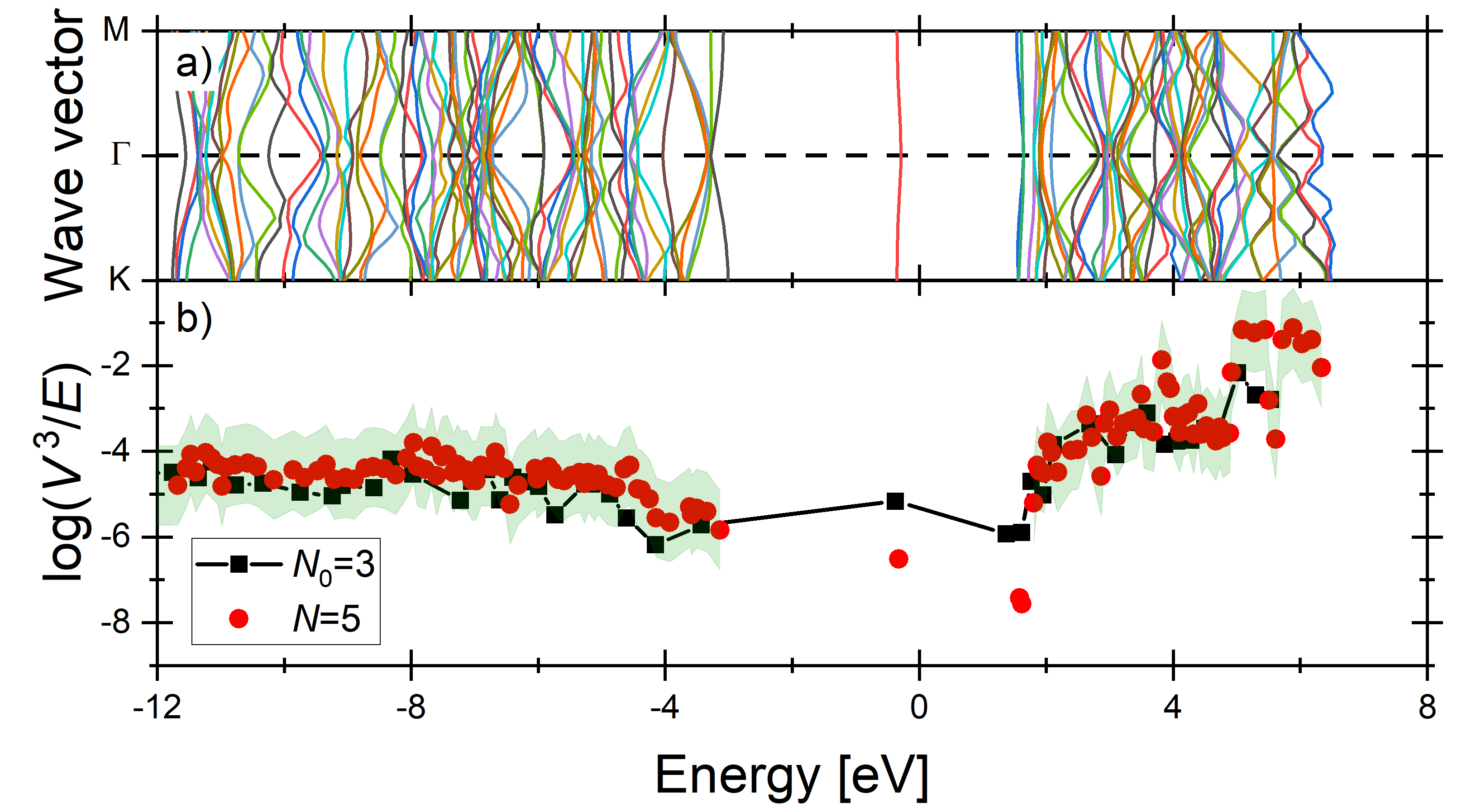}
	\centering
	\caption{Confinement analysis applied to electronic waves in a 2D hexagonal BN with nitrogen deficiency, $j=1$. 
         Zero energy is set at the Fermi level. 
         (a) Band structure of the $N=5$ supercell, flipped to have an energy abscissa.
         Different bands are shown with different color for clarity. 
         (b) Scaling analysis of confinement. 
         Every band is represented by a point.
         These results confirm the three point-confined bands around the Fermi level.
         However, compared to Fig.~\ref{main-fig:DFTconfinement}, several other bands appear slightly under the reference line.
         We attribute this effect to be most likely caused by the sub-leading order contributions in the scaling and would be remedied for larger supercell sizes.
         This conclusion is enhanced by the fact that the standard deviation for these bands (green area) is higher than their separation from the reference line.}
	\label{DFTj1}
\end{figure}

For completeness, we include the plot for $j=1$, which was omitted in the main text, in Fig.~\ref{DFTj1}.
These results confirm the three point-confined bands ($c=2$ in a $D=2$ system) around the Fermi level.
At the same time, some other bands of the $N=5$ supercell seem to be slightly under the reference values.
Naively, this could indicate $c=1$ confinement dimensionality for those bands, however, in this case we attribute this effect to be most likely caused by the sub-leading order effects in the scaling.
This would be remedied by using larger supercell sizes $N$ and $N_0$.
This conclusion is confirmed by looking at the standard deviation of the $c\ne2$ bands for the $N=5$ supercell (green area in Fig.~\ref{DFTj1}(b)), which is higher than the separation of the questionable bands from their reference line, in each case.
Fig.~\ref{DFTj1} exemplifies that our framework can offer rigorous and cost-effective scaling analysis, but, especially for small supercell sizes, its results must still be critically evaluated to avoid the sub-leading order effects of scaling.

\section{Photonic band computation}
\begin{figure}[H]
	\includegraphics[width=\linewidth]{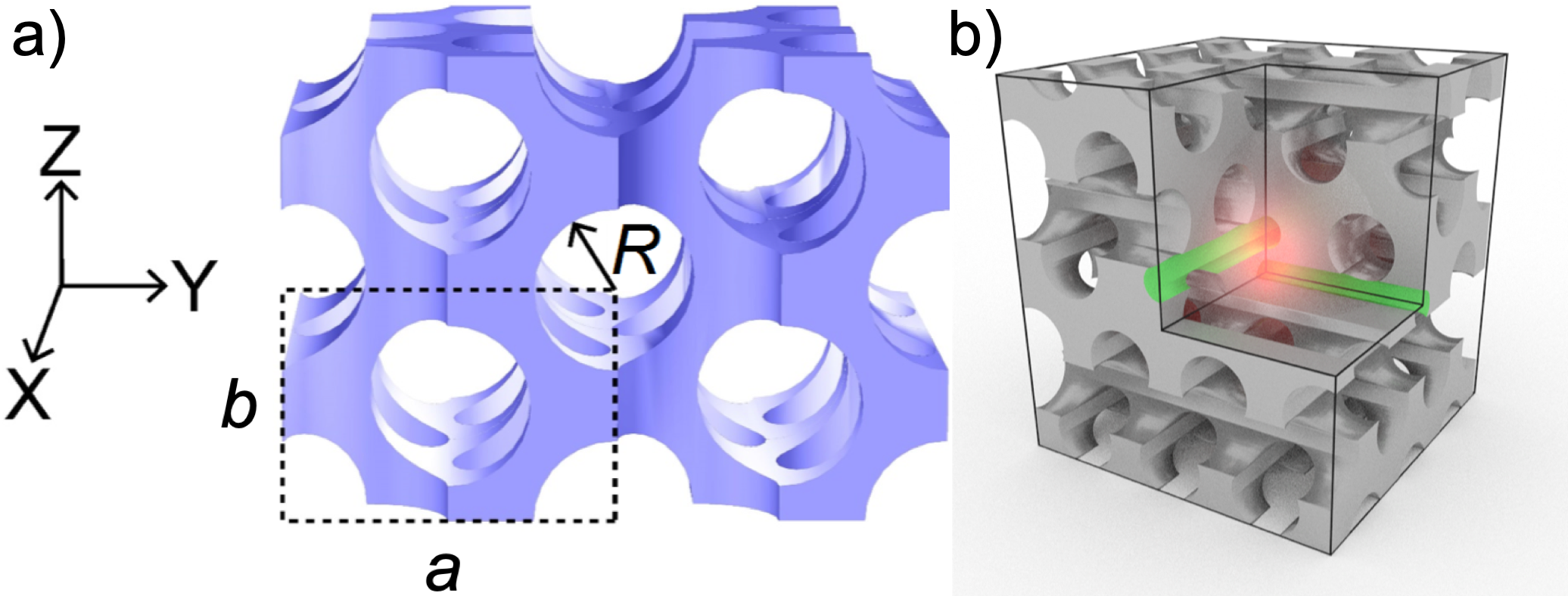}
	\centering
	\caption{\label{fig:struc}(a) Structure of an unperturbed inverse woodpile photonic crystal. We use a tetragonal unit cell with lattice parameters $a$ in the $y$ direction and $b=a/\sqrt{2}$ in the $x$ and $z$ directions. The pore radius is denoted by $R$. The figure shows an $N=2$ supercell, with one unit cell designated by the dashed border. (b) Design of the defect. The radius $R^\pr$ of two proximate defect pores (shown in green) is altered. At their intersection a region with excess of one material is created that serves as a point defect (red glow).}
	\label{photonicstructure}
\end{figure}

In the main text we analyze light confinement in a 3D inverse woodpile photonic band gap superlattice with two proximate linear defects.
The photonic band gap crystal consists of bulk silicon ($\eps = 12.1$, see Ref.~\cite{Hillebrand2003J.Appl.Phys.}), in which nanopores of radius $R$ filled with air ($\eps=1$) are etched \cite{Ho1994SolidStateCommun.,Leistikow2011Phys.Rev.Lett.}.
We model the unperturbed crystal using a tetragonal unit cell with lattice parameters $a$ in the $y$ direction and $b=\nicefrac{a}{\sqrt{2}}$ in the $x$ and $z$ directions.
This unperturbed structure is depicted in Fig.~\ref{photonicstructure}(a).

A defect can be incorporated in the structure by altering the radius $R^\pr$ of two proximate nanopores, as illustrated in Fig.~\ref{photonicstructure}(b).
In our example from the main text we use the unperturbed pores of the radius $R = 0.24a$ and the defect pores with the increased radius $R^\pr = 0.35 a > R$.
This results in creation of a region with excess air at the intersection of the two defect pores, serving as a point defect.
We include one defect of this type per supercell.
Such a defect results in splitting of the defect bands from the bottom edge of the band gap, in analogy with acceptor-doped lattices in solid state physics.

We investigate the confinement of $N=4$ supercell and use $N_0=2$ as a reference.
The spectrum of the investigated cavity superlattice is computed by the plane-wave expansion method using the well-known MPB code~\cite{Johnson2001Opt.Express} and is depicted in the main text in Fig.~\ref{main-fig:photonicconfinement}(a).
For the band structure, we follow the high-symmetry path of the tetragonal unit cell according to Ref.~\cite{Setyawan2010Comput.Mater.Sci.}.
In order to use unified notation, we re-labeled our high-symmetry points and axes in Fig.~\ref{main-fig:photonicconfinement}(a) according to Ref.~\cite{Setyawan2010Comput.Mater.Sci.}, which assumes that the two equal lattice constants are in the $x$ and $y$ directions.

\begin{figure}
	\includegraphics[width=\linewidth]{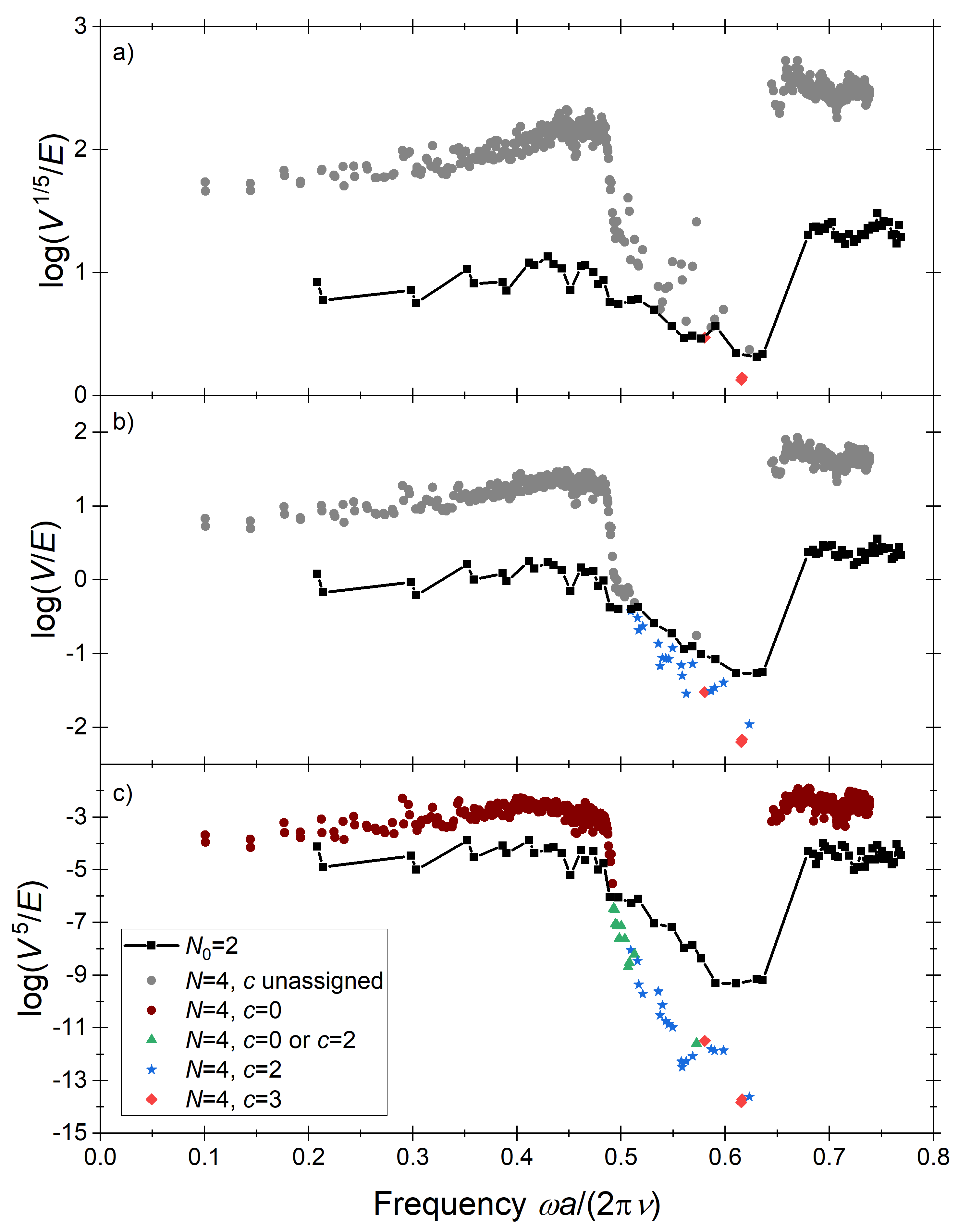}
	\centering
	\caption{Step-by-step confinement analysis of bands for a 3D inverse woodpile photonic crystal with two proximate line defects.
	Every band is represented by a point.
	The values of the auxiliary power $\alpha$ for each $j$ are (a) $j=3, \alpha=1/5$, (b) $j=2, \alpha=1$, (c) $j=1, \alpha=5$, as per Table~\ref{main-table:kappa}.}
	\label{fig:confanalysis}
\end{figure}

We perform scaling analysis to identify the confinement dimensionality $c$ of each band, as described by our theory introduced in this Letter.
Each step of the analysis for this $D=3$-dimensional system is described in Fig.~\ref{fig:confanalysis}.
We choose the auxiliary power $\alpha$ based on Table~\ref{main-table:kappa}.

Fig.~\ref{fig:confanalysis}(a) describes $j=3$, with $\alpha=\nicefrac{1}{5}$.
Out of the whole spectrum of unidentified bands, three bands around $\tilde\omega=0.6$ decrease in value of $\log(V^{\nicefrac{1}{5}}/E)$ with respect to the reference supercell.
We thus identify these bands as having $c=3$ confinement dimensionality.
This is in stark contrast with the statement made in Ref.~\cite{Woldering2014Phys.Rev.B} that no point-confined bands exist in acceptor-like 3D inverse woodpile crystals.

Fig.~\ref{fig:confanalysis}(b) describes $j=2$, with $\alpha=1$.
In this case, several additional bands between roughly $\tilde\omega = 0.5$ and $\tilde\omega = 0.6$ drop below the values of the reference supercell.
These newly separated bands are therefore identified as having $c=2$ confinement dimensionality.
This marks the first time such $c=2$ bands are distinguished from fully extended bands in a 3D inverse woodpile photonic structure.

Fig.~\ref{fig:confanalysis}(c) describes $j=1$, with $\alpha=5$.
There are several more bands dropping below the reference line.
According to our framework, this would mean that they are plane-confined, \ie , having $c=1$.
Nevertheless, we know that our defect geometry is linear and therefore cannot support plane-confined bands.
These bands must thus either have $c=2$ or $c=0$.

The misidentification for bands that should have $c=2$ is due to the fact that the size of the (reference) supercell is still below the localization length in the plane where the linear defects are positioned.
However, the scaling Eq.~\eq{main-eq:scaling} holds only for supercell sizes larger than the localization length.
Therefore it is understandable that our method "thinks" that these bands are plane-confined.
Simply put, for the given supercell size, the information about confinement of these bands is not yet available in the calculated energy density and thus cannot be resolved by any means, unless implementing some additional knowledge, for example about the defect geometry.

The misidentified bands that should have $c=0$ are caused by the fact that for the given supercell size the effects of the sub-leading order terms on Eq.~\eq{main-eq:scaling} are still too large.
This is an issue of the convergence of our technique and it would also be corrected for larger supercell sizes.
In this case, it could be possible to devise some alternative way to improve the convergence since, unlike in the case of the previous paragraph, the information about the confinement may be available in the calculated energy density data - this remains, however, beyond the scope of this Letter.

Lastly, all the bands remaining above the reference values in Fig.~\ref{fig:confanalysis}(c) are identified as extended, \ie , with $c=0$.

Our method is especially powerful for experimentally interesting supercells of moderate size.
These supercell sizes are, however, in a range where possible inaccuracies due to finite-size scaling can occur and therefore, naturally, the results need to be critically evaluated.
The presented photonic example further illustrates the strength and novelty of our method. 
Our framework has enabled the discovery of 3D-confined acceptor-like bands previously thought non-existent and, for the first time ever in an inverse woodpile structure, distinguished 2D-confined bands from the extended ones.
It is therefore clear that our technique enables direct access to previously inaccessible confinement information.

\bibliography{Cavity_superlattice_scaling} 

\makeatletter\@input{xx.tex}\makeatother 